\documentclass[journal,12pt,draftclsnofoot,onecolumn]{IEEEtran}

\usepackage{cite}
\usepackage{amsmath,amssymb,amsfonts}
\usepackage{algorithm}  
\usepackage{algpseudocode}  
\usepackage{algpseudocode}
\usepackage{graphicx}
\usepackage{epstopdf}
\usepackage{textcomp}
\usepackage{xcolor}
\usepackage{bm}
\usepackage{setspace}

\ifCLASSINFOpdf
\else
\fi
\begin{document}
	
%
\title{Robust Beamforming Design and Time Allocation for IRS-assisted Wireless Powered Communication Networks}
%
%
%
\author{Zhendong~Li,~Wen~Chen,~\IEEEmembership{Senior~Member,~IEEE},~Qingqing~Wu,~\IEEEmembership{Member,~IEEE},~Huanqing~Cao,~Kunlun~Wang,~\IEEEmembership{Member,~IEEE},~and~Jun~Li,~\IEEEmembership{Senior~Member,~IEEE}
\thanks{Z. Li, W. Chen and H. Cao are with the Department of Electronic Engineering, Shanghai Jiao Tong University, Shanghai 200240, China (e-mail: lizhendong@sjtu.edu.cn; wenchen@sjtu.edu.cn; caohuanqing@sjtu.edu.cn).}
\thanks{Q. Wu is with the State Key Laboratory of Internet of Things for Smart City, University of Macau, Macau 999078, China, and also with the National Mobile Communications Research Laboratory, Southeast University, Nanjing 210096, China (e-mail: qingqingwu@um.edu.mo). }
\thanks{K. Wang is with the School of Communication and Electronic Engineering, East China Normal University, Shanghai 200241, China (e-mail: klwang@cee.ecnu.edu.cn).}
\thanks{J. Li is with the School of Electronic and Optical Engineering, Nanjing University of Science Technology, Nanjing 210094, China (email: jun.li@njust.edu.cn). }}
\maketitle

\begin{abstract}
	In this paper, a novel intelligent reflecting surface (IRS)-assisted wireless powered communication network (WPCN) architecture is proposed for power-constrained Internet-of-Things (IoT) smart devices, where IRS is exploited to improve the performance	of WPCN under imperfect channel state information (CSI). We formulate a hybrid access point (HAP) transmit energy minimization problem by jointly optimizing time allocation, HAP energy beamforming, receiving beamforming, user transmit power allocation, IRS energy reflection coefficient and information reflection coefficient under the imperfect CSI and non-linear energy harvesting model. On account of the high coupling of optimization variables, the formulated problem is a non-convex optimization problem that is difficult to solve directly. To address the above-mentioned challenging problem, alternating optimization (AO) technique is applied to decouple the optimization variables to solve the problem. Specifically, through AO, time allocation, HAP energy beamforming, receiving beamforming, user transmit power allocation, IRS energy reflection coefficient and information reflection coefficient are divided into three sub-problems to be solved alternately. The difference-of-convex (DC) programming is used to solve the non-convex rank-one constraint in solving IRS energy reflection coefficient and information reflection coefficient. Numerical simulations verify the superiority of the proposed optimization algorithm in decreasing HAP transmit energy compared with other benchmark schemes.
\end{abstract}


%
\IEEEpeerreviewmaketitle

\section{Introduction}
%
%
%
%
\IEEEPARstart{I}{n} recent years, the number of smart devices has also shown a blowout growth along with the vigorous development of Internet-of-Things (IoT) \cite{9060972,7320954}. There will be more than 28 billion devices connected to wireless networks in 2022 according to Cisco's forecast \cite{forecast2019cisco}. Nowadays, there are two main challenges in the actual deployment of IoT networks. First, smart devices are usually powered by batteries, which have a short life cycle, and the cost of battery replacement and redeployment is relatively high. In addition, the interference problem caused by the large-scale access of smart devices will greatly reduce the system capacity. Therefore, a novel paradigm that can efficiently satisfy the IoT network needs to be proposed.

As an effective wireless energy network mode, wireless powered communication network (WPCN) was first proposed in \cite{6678102}. Ju \emph{et al.} first proposed the ``harvest-then-transmit" protocol, i.e., hybrid access point (HAP) can be used to provide users with continuous downlink energy services, while collecting uplink transmission information generated by users based on time division multiple access (TDMA) \cite{6678102}. This protocol can be better applied to most IoT scenarios, where HAP provides energy services for energy-constrained smart devices, and then the devices perform data transmission. In recent years, research on WPCN in IoT networks has also received extensive attention from academia \cite{9316713,9149891,8982086,8489918,9084128,7332956}. Aiming at the energy-constrainted nodes in the IoT network,  Song \emph{et al.} proposed a new iterative algorithm for cluster-specific beamforming design using the WPCN framework, and then proposed an algorithm for maximizing total throughput \cite{9316713}. In the WPCN-assisted IoT network, Nguyen \emph{et al.} considered users to access by orthogonal frequency division multiple access (OFDMA), and jointly optimized the duration of energy harvesting (EH), subcarriers and power allocation to maximize the system energy efficiency (EE) \cite{9149891}. For self-sustaining wireless communication systems in IoT networks, channel allocation, time resource and power resource allocation were optimized to maximize uplink weighted sum-rate \cite{9084128}. However, previous research on WPCN in IoT networks still faces two main challenges: one is low energy efficiency and small coverage, and the other is the interference caused by large-scale access of massive devices. For the first challenge, it is mainly because that the distance-related channel attenuation will greatly affect the energy harvesting of smart devices. Therefore, previous solutions have focused on reducing the distance between the device and access point (AP), so that the transmission between smart device and the AP has a good line-of-sight (LoS) link. However, it greatly limits the energy transmission efficiency of the AP and the coverage of the network. Therefore, it is urgent to adopt new and cost-effective solutions to improve the performance of WPCN in IoT networks.

Intelligent reflecting surface (IRS), also known as reconfigurable intelligent surface (RIS), has been widely discussed in the industry and academia as an emerging technology \cite{9167258,8811733,8741198,9048622}. By changing the amplitude and phase shift of incident electromagnetic (EM) waves, IRS can reconstruct the wireless propagation channel between transceivers. Then, the network capacity, energy efficiency and coverage can be improved with low cost. The IRS is an array composed of a large number of passive reflective units, and its deployment is more convenient, such as on a wall or a building. Since the IRS is passive, it only reflects the incident signal compared with the relay and does not process the signal, so it does not introduce additional noise \cite{gong2019towards,zhao2019survey}. In addition, since it is not equipped with a wireless radio frequency (RF) link, compared with multiple-input multiple-output (MIMO) system, it can greatly reduce cost and power consumption \cite{9117136}. The above-mentioned significant advantages make IRS have a very broad application prospect in future communication networks.

Due to the many advantages of IRS, the assistance of IRS can greatly improve the energy efficiency and coverage of WPCN in IoT networks. In order to better illustrate the superiority of IRS, some scholars have studied the WPCN network assisted by IRS \cite{9214497,9298890,9003222,9485102,9531372,9388935,9509394}. Lyu \emph{et al.} proposed a hybrid relay scheme for self-sustaining IRS in WPCN to simultaneously improve the performance of downlink energy transmission from HAP to multiple users and the performance of uplink information transmission from users to HAP \cite{9214497}. Zheng \emph{et al.} jointly optimized HAP active beamforming and IRS passive beamforming and user power for multi-user multiple-input single-output (MISO) networks to study the weighted sum-rate maximization of IRS-assisted WPCN \cite{9298890}. \cite{9003222} considered the cooperation of users assisted by IRS in the WPCN, where two users harvest wireless energy and transmit the information to the HAP. For a downlink RIS-assisted WPCN, Xu \emph{et al.} proposed a joint radio resource and passive beamforming optimization algorithm to improve system energy efficiency (EE) \cite{9485102}. However, the existing IRS-assisted WPCN research is still in its infancy and most of the existing works consider the linear energy harvesting model. While this assumption is not practical, so more practical nonlinear energy harvesting models need to be considered, which also greatly stimulated this work.

The second challenge mentioned above is that with the proliferation of smart devices, the problem of interference between devices caused by large-access will become more serious, and system capacity will be limited. Beamforming is a very effective solution to deal with the interference problem caused by the access of multiple users and increase the system capacity \cite{4350296}. For beamforming technology, channel state information (CSI) is critical, which is usually obtained through channel estimation. However, in practical systems, the uncertainty of channel estimation brings greater challenge to the beamforming technology. Take the previously mentioned IRS-assisted WPCN as an example, HAP needs to provide energy beam services for smart devices based on CSI, and then can receive signals transmitted by users. The performance of both depends largely on channel estimation. Previous researches on IRS-assisted WPCN are mostly aimed at the single-antenna case \cite{9214497,9400380}, or the case of perfect CSI \cite{9214497,9298890,9003222}, which will reduce the performance of the IRS-assisted WPCN in the practical IoT networks. Therefore, in order to solve the aforementioned challenges, we consider beamforming in the case of imperfect CSI, which is novel and closer to the practical system.

In summary, for the large-scale access of power-constrained devices in IoT networks, the increased demands for network energy efficiency, coverage, and system capacity have greatly spurred the research on robust IRS-assisted WPCN. To our best knowledge, time allocation and robust active and passive beamforming design in IRS-assisted WPCN has not been investigated in the literature. In this paper, considering that HAP cannot obtain perfect CSI, we minimize HAP transmit energy consumption by jointly optimizing time allocation, HAP energy beamforming in downlink (DL), IRS energy reflection coefficient, HAP receiving beamforming in uplink (UL), IRS information reflection coefficient, and user transmit power allocation in UL. Since the HAP cannot obtain perfect CSI and high coupling of optimization variables, it is not easy to deal with this formulated problem. Consequently, it is necessary to design an effective joint optimization algorithm for the IRS-assisted WPCN in IoT networks, which can solve HAP transmit energy minimization problem. The main contributions of this paper are as follows:
\begin{itemize}
\item We propose a novel IRS-assisted WPCN framework in IoT networks, where the ``harvest-then-transmit" protocol is adopted. Specifically, the HAP first provides DL energy beamforming for multiple users with the assistance of IRS, and then the users use the harvested energy for UL information transmission with the assistance of IRS. The IRS can change the wireless channel to improve the energy transmission efficiency and coverage of WPCN. In addition, our proposed framework adopts a non-linear energy harvesting model and considers the case of imperfect CSI, which thus makes it more practical than existing works. We formulate the HAP transmit energy minimization problem for joint optimization of time allocation, HAP energy beamforming, IRS energy reflection coefficient in DL, HAP receiving beamforming, IRS information reflection coefficient in UL, and user transmit power allocate in UL. In view of the coupling of optimization variables, the minimization problem is non-convex, and it is not easy to obtain the solution of the problem.

\item To address the formulated HAP transmit energy minimization problem, we first transform the problem and then use alternating optimization (AO) technique to divide the transformed problem into three sub-problems. More specifically, in the first sub-problem, given time allocation and IRS reflection coefficient, we propose optimization algorithms for the HAP energy beamforming, receiving beamforming, and user transmit power allocation. In the second sub-problem, we propose an optimization algorithm for IRS energy reflection coefficient and information reflection coefficient based on the difference-of-convex (DC) programming. For the third sub-problem, we solve a linear programming problem (LP) to obtain time allocation scheme. Finally, the three sub-problems are alternately optimized until the overall problem converges.

\item Numerical simulation results confirm the performance advantage of the proposed algorithm compared with other benchmarks, i.e., it can significantly reduce the HAP transmit energy. For IRS-assisted WPCN, the HAP transmit energy is lower than that of networks without IRS assistance. Meanwhile, the more IRS reflection elements, the smaller the transmit energy required. The applied DC programming can also solve the problem of non-convex rank-one constraint in the second sub-problem, and its performance is better than the commonly used semidefinite relaxation (SDR). In addition, the performance of our proposed algorithm is close to that in the case of perfect CSI, but it is more realistic and has better robustness.
\end{itemize}


\textit{Notations:} In this paper, scalars are denoted by lower-case letters. Vectors and matrices are respectively represented by bold lower-case letters and bold upper-case letters. The absolute value of a complex-valued scalar $x$ can be denoted by $\left| {x} \right|$, and the Euclidean norm of a complex-valued vector $\bf{x}$ can be denoted by $\left\| {\bf{x}} \right\|$. In addition, $\rm{tr(\bf{X})}$, $\rm{rank(\bf{X})}$, ${\bf{X}}^H$, ${\bf{X}}_{m,n}$ and $\left\| {\bf{X}} \right\|$ denote trace, rank, conjugate transpose, $m,n$-th entry and matrix norm of a square matrix $\bf{X}$, respectively, while ${\bf{X}} \succeq 0$ represents the square matrix $\bf{X}$ is a positive semidefinite matrix. Similarly, $\rm{rank(\bf{A})}$, ${\bf{A}}^H$, ${\bf{A}}_{m,n}$ and $\left\| {\bf{A}} \right\|$ also denote rank, conjugate transpose, $m,n$-th entry and matrix norm of a general matrix $\bf{A}$, respectively. ${\mathbb{C}^{M \times N}}$ represents the space of ${M \times N}$ complex matrix. ${\bf{I}}_N$ represents an identity matrix of size ${N \times N}$. $j$ represents the imaginary unit, i.e., $j^2=-1$. $\mathbb{E}\left\{  \cdot  \right\}$ is expectation operator. Finally, the distribution of a circularly symmetric complex Gaussian (CSCG) random vector with mean $\mu$ and covariance matrix $\bf{C}$ can be expressed as $ {\cal C}{\cal N}\left( {\mu,\bf{C}} \right)$, and $\sim$ denotes `distributed as'.
\section{System model and problem formulation}
\subsection{System model}
In this section, we first introduce the system model of IRS-assisted WPCN in IoT networks, as shown in Fig. 1, which includes an HAP with $N$ antennas, an IRS with $M$ reflection elements deployed on a building, and $K$ users (i.e. IoT devices) with a single antenna. The HAP with constant power can be used to coordinate the energy transmission and information transmission of a group of users. In WPCN, HAP means that the energy station and the information station are hybrid, and it can broadcast energy signals to users via the DL as well as receive user information via the UL. The antennas of the HAP and the array of the IRS are both distributed in a uniform linear array (ULA)\footnote{Note that if HAP and IRS use uniform planar array (UPA) instead of ULA for channel modeling, the optimization algorithm proposed in this paper can still work well.}. Let ${\bf{E}} = \sqrt \iota  {\rm{diag}}\left( {{e_1},...,{e_M}} \right) \in {{\mathbb{C}}^{M \times M}}$ denote the energy reflection coefficient matrix of the IRS in the DL, and let ${\bf{Q}} = \sqrt \varpi  {\rm{diag}}\left( {{q_1},...,{q_M}} \right) \in {{\mathbb{C}}^{M \times M}}$ denote the information reflection coefficient matrix of the IRS in the UL, which has the following constraints,
\begin{equation}
	{\left| {{e_m}} \right|^2} = 1,\forall m,
\end{equation}
\begin{equation}
	{\left| {{q_m}} \right|^2} = 1,\forall m.
\end{equation}
In order to maximize the strength of the reflective signal, the amplitude of each element of the IRS is usually set to 1 \cite{8811733}. Meanwhile, for the IRS structure, the phase shift is mainly controlled by the conduction of the diode, so it is more difficult to adjust its amplitude, which will also bring certain challenges to the design of the IRS \cite{9133266}. Hence, we use an IRS with a fixed amplitude of 1, i.e., $\iota  = 1$ and $\varpi  = 1$. In addition, due to severe path loss, the signal reflected twice or more by the IRS can be ignored. We assume that the phase shift of the IRS can be calculated by the HAP and sent to the IRS controller through the feedback channel. For DL wireless energy transmission, the channel gain from HAP to IRS, from IRS to the $k$-th user, and from HAP to the $k$-th user can be expressed as ${{\bf{H}}_{d,r}} \in {{\mathbb{C}}^{M \times N}}$, ${\bf{h}}_{r,k}^H \in {{\mathbb{C}}^{1 \times M}}$, and ${\bf{h}}_{d,k}^H \in {{\mathbb{C}}^{1 \times N}}$, respectively. We can use the existing channel estimation algorithm to estimate the DL channel \cite{8879620,9328501,9398559}, and then according to the channel reciprocity, the UL CSI can be obtained. Hence, for UL wireless information transmission, the channel gain from the IRS to the HAP, from the $k$-th user to the IRS, and from the $k$-th user to the HAP can also be expressed as ${{\bf{H}}_{d,r}} \in {{\mathbb{C}}^{M \times N}}$, ${\bf{h}}_{r,k}^H \in {{\mathbb{C}}^{1 \times M}}$, and ${\bf{h}}_{d,k}^H \in {{\mathbb{C}}^{1 \times N}}$. The channel consists of a direct channel and a cascaded channel, named HAP-user channel and HAP-IRS-user channel, respectively. We assume that all channels are quasi-static flat fading, i.e., in each transmission time duration $T$, ${{\bf{H}}_{d,r}}$, ${\bf{h}}_{r,k}^H$ and ${\bf{h}}_{d,k}^H$ are constant.

\begin{figure}
	\centerline{\includegraphics[width=10cm]{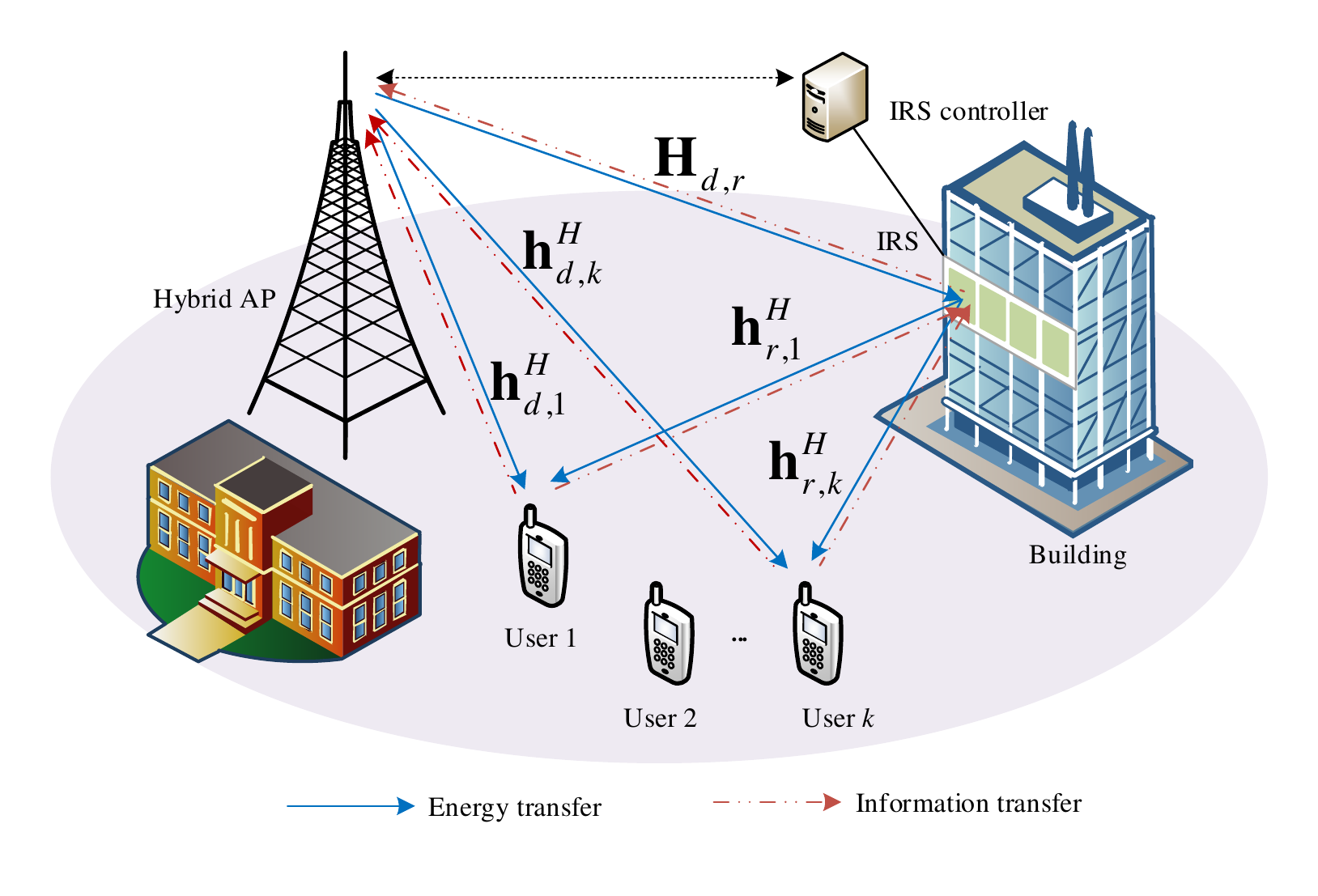}}
	\caption{IRS-assisted WPCN in IoT networks.}
	\label{Fig1}
\end{figure}

Although the channel between the HAP and the user may be blocked, the wireless channel still has a lot of scattering, so the channel between the HAP and the user can be modeled as Rayleigh fading, denoted as ${\bf{g}}_{d,k}^H \in {{\mathbb{C}}^{1 \times N}}$. We assume that each element of ${\bf{g}}_{d,k}^H$ is independent and identically distributed (i.i.d) CSCG random variable with zero mean and unit variance \cite{8811733,9167258}. Therefore, the channel gain between HAP and the $k$-th user is
\begin{equation}
	{\bf{h}}_{d,k}^H = \sqrt {{C_0}{{\left( {\frac{{{d_{d,k}}}}{{{D_0}}}} \right)}^{ - \alpha }}} {\bf{g}}_{d,k}^H,
\end{equation}
where ${C_0}$ is the path loss at the reference distance ${D_0} = 1{\rm{m}}$, ${d_{d,k}}$ is the distance between the HAP and the $k$-th user, and $\alpha $ is the path loss exponent between the HAP and the $k$-th user. In addition, the channel between HAP and IRS and the channel between IRS and users have line-of-sight (LoS) components, so we model them as Rician channels, which can be given by
\begin{equation}
	{{\bf{\bar H}}_{d,r}} = \sqrt {\frac{\kappa }{{1 + \kappa }}} {\bf{H}}_{d,r}^{{\rm{LoS}}} + \sqrt {\frac{1}{{1 + \kappa }}} {\bf{H}}_{d,r}^{{\rm{NLoS}}},
\end{equation}
and
\begin{equation}
	{\bf{\bar h}}_{r,k}^H = \sqrt {\frac{\vartheta }{{1 + \vartheta }}} {\bf{h}}_{r,k}^{{\rm{LoS}}} + \sqrt {\frac{1}{{1 + \vartheta }}} {\bf{h}}_{r,k}^{{\rm{NLoS}}},
\end{equation}
where $\kappa $ and $\vartheta $ respectively represent the Rice factor of the corresponding channel, ${\bf{H}}_{d,r}^{{\rm{LoS}}}$ and ${\bf{h}}_{r,k}^{{\rm{LoS}}}$ respectively represent the LoS component of the corresponding channel. ${\bf{H}}_{d,r}^{{\rm{NLoS}}}$ and ${\bf{h}}_{r,k}^{{\rm{NLoS}}}$ respectively represent the non-line-of-sight (NLoS) component of the corresponding channel, and each of their elements is i.i.d CSCG random variable with zero mean and unit variance \cite{8811733,9167258}.

The LoS component is represented by the array response of ULA. The array response of $N$ elements ULA can be given by
\begin{equation}
	{{\bf{a}}_N}\left( \theta  \right) = \left[ {1,{e^{ - j2\pi \frac{d}{\lambda }\sin \theta }},...,{e^{ - j2\pi \left( {N - 1} \right)\frac{d}{\lambda }\sin \theta }}} \right],
\end{equation}
where $\theta$ represents the angle-of-arrival (AoA) or the angle-of-departure (AoD) of the signal\footnote{Note that the angle of arrival of all elements is the same, but the antenna response is different.}, $\lambda $ denotes the carrier wavelength and $d$ represents the spacing between adjacent antenna elements. Therefore, the LoS component ${\bf{H}}_{d,r}^{{\rm{LoS}}}$ and ${\bf{h}}_{r,k}^{{\rm{LoS}}}$ can be given by
\begin{equation}
	{\bf{H}}_{d,r}^{{\rm{LoS}}} = {\bf{a}}_M^H\left( {{\theta _{{\rm{AoA}},{\rm{1}}}}} \right){{\bf{a}}_N}\left( {{\theta _{{\rm{AoD}},{\rm{1}}}}} \right),
\end{equation}
and
\begin{equation}
	{\bf{h}}_{r,k}^{{\rm{LoS}}} = {{\bf{a}}_M}\left( {{\theta _{{\rm{AoD}},{\rm{2}}}}} \right),
\end{equation}
where ${{\theta _{{\rm{AoA,1}}}}}$ is the AoA to the ULA at IRS, and ${{\theta _{{\rm{AoD,1}}}}}$ is the AoD from the ULA at HAP. ${{\theta _{{\rm{AoD,2}}}}}$ is the AoD from the ULA at IRS.  Therefore, The channel gain between HAP and IRS can be expressed as
\begin{equation}
	{{\bf{H}}_{d,r}} = \sqrt {{C_0}{{\left( {\frac{{{d_{d,r}}}}{{{D_0}}}} \right)}^{ - \beta }}} {{\bf{\bar H}}_{d,r}},
\end{equation}
and the channel gain between IRS and the $k$-th user can be given by
\begin{equation}
	{\bf{h}}_{r,k}^H = \sqrt {{C_0}{{\left( {\frac{{{d_{r,k}}}}{{{D_0}}}} \right)}^{ - o}}} {\bf{\bar h}}_{r,k}^H,
\end{equation}
where ${{d_{d,r}}}$ and ${{d_{r,k}}}$ represent the distance from HAP to IRS and the distance from IRS to the $k$-th user, respectively. $\beta$ and $o$ respectively represent path loss exponent from HAP to IRS and IRS to the $k$-th user. 

In this paper, we consider that all users do not have traditional energy supply (e.g., power supply, etc.), so they need to harvest energy from the HAP transmission signal in the DL. We assume that each user is equipped with an energy storage device to store energy obtained from radio frequency signals for wireless information transmission. We adopt the ``harvest-then-transmit" protocol. Each transmission time duration $T$ is divided into two parts\footnote{We assume that in each transmission time duration $T$, users are static, i.e., the location of users does not change.}, as shown in Fig. 2. The HAP broadcasts energy signals to all users to provide users with continuous energy in $\tau T\left( {0 < \tau  < 1} \right)$. Then, all users use the energy they harvest in DL to transmit their independent information to the multi-antenna HAP in UL by space division multiple access (SDMA) in $\left( {1 - \tau } \right)T$. We normalize time to $T = 1$ without loss of generality in this paper.
\begin{figure}
	\centerline{\includegraphics[width=10cm]{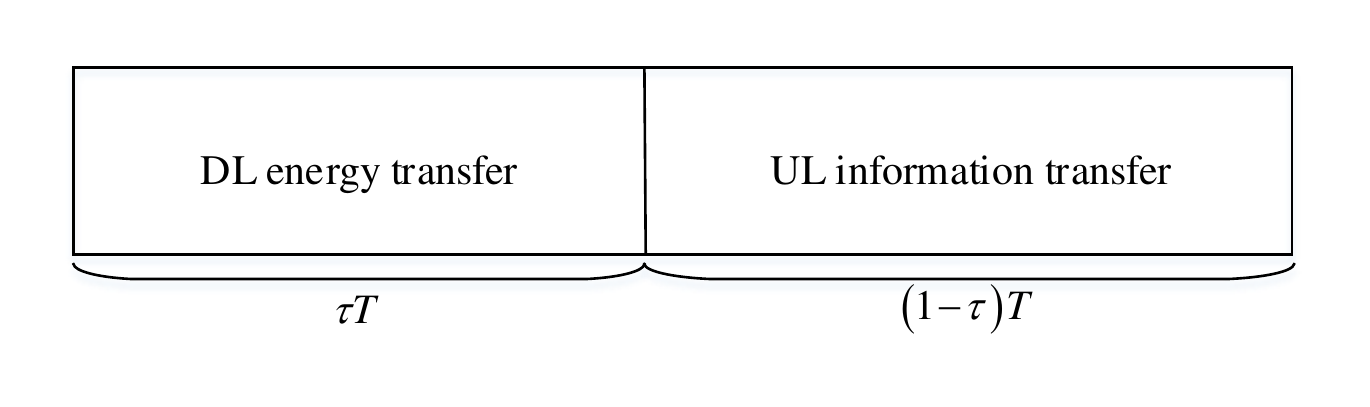}}
	\caption{``Harvest-then-transmit" protocol.}
	\label{Fig2}
\end{figure}

More specifically, in DL, the HAP sends $L$ energy beams to all users, and $L$ can be any integer not exceeding $N$. The baseband transmission signal can be expressed as
\begin{equation}
	{{\bf{x}}_0} = \sum\limits_{l = 1}^L {{{\bf{v}}_l}s_l^{{\rm{dl}}}} ,
\end{equation}
where $s_l^{{\rm{dl}}}$ denotes the energy signal, which is assumed to be i.i.d CSCG random variable with zero mean and unit variance, i.e., $s_l^{{\rm{dl}}} \sim {\cal C}{\cal N}\left( {0,1} \right)$. ${{\bf{v}}_l} \in {\mathbb{C}^{N \times 1}}$ represents the $l$-th energy beam, and the HAP transmission power in DL can be denoted by ${\rm \mathbb{E}}\left\{ {{{\left\| {{{\bf{x}}_0}} \right\|}^2}} \right\} = \sum\limits_{l = 1}^L {{{\left\| {{{\bf{v}}_l}} \right\|}^2}} $. Let $P_{\max }$ be the maximum transmit power of HAP\footnote{In this paper, we mainly consider the transmit energy minimization, the power consumption required for channel estimation and the circuit power consumption of the IRS controller are temporarily ignored. They can be considered when we solve other optimization problems, e.g., the problem of maximizing energy harvested by users.}, we have
\begin{equation}
	\sum\limits_{l = 1}^L {{{\left\| {{{\bf{v}}_l}} \right\|}^2}}  \le {P_{\max }}.
\end{equation}
In the DL, through the HAP-user channel and HAP-IRS-user channel, the energy signal obtained by the $k$-th user is given by
\begin{equation}
	{y_k} = \left( {{\bf{h}}_{r,k}^H{\bf{E}}{{\bf{H}}_{d,r}} + {\bf{h}}_{d,k}^H} \right){{\bf{x}}_0} = \left( {{\bf{h}}_{r,k}^H{\bf{E}}{{\bf{H}}_{d,r}} + {\bf{h}}_{d,k}^H} \right)\sum\limits_{l = 1}^L {{{\bf{v}}_l}s_l^{{\rm{dl}}}} .
\end{equation}
Let ${\bf{e}} = {\left[ {{e_1},...,{e_M}} \right]^T} \in {\mathbb{C}^{M \times 1}}$ and ${{\bf{G}}_k} = {\rm{diag}}\left( {{\bf{h}}_{r,k}^H} \right){{\bf{H}}_{d,r}}$. Then the energy signal obtained by the $k$-th user is further given by
\begin{equation}
	{y_k} = \left( {{{\bf{e}}^H}{{\bf{G}}_k} + {\bf{h}}_{d,k}^H} \right)\sum\limits_{l = 1}^L {{{\bf{v}}_l}s_l^{{\rm{dl}}}} ,\forall k.
\end{equation}
Since the HAP in the DL adopts a broadcast mode, all $L$ energy beams can be received for the $k$-th user. Therefore, the received power in the DL for the $k$-th user can be given
\begin{equation}
	{P_k} = {\mathbb{E}}\left\{ {{{\left| {{y_k}} \right|}^2}} \right\} = {\sum\limits_{l = 1}^L {\left| {\left( {{{\bf{e}}^H}{{\bf{G}}_k} + {\bf{h}}_{d,k}^H} \right){{\bf{v}}_l}} \right|} ^2},\forall k.
\end{equation}
We adopt the more practical non-linear energy harvesting model in this paper. Accordingly, the harvested power of the $k$-th user can be expressed as
\begin{equation}
	\Xi \left( {{P_k}} \right) = \left( {\frac{{{\xi _k}}}{{{X_k}\left( {1 + \exp \left( { - {a_k}\left( {{P_k} - {b_k}} \right)} \right)} \right)}} - {Y_k}} \right),\forall k,
\end{equation}
where $\xi _k$ is the maximum power that the $k$-th user can harvest. ${a_k}$ and ${b_k}$ denote parameters related to specific circuit specifications. Herein, ${X_k} = {{\exp \left( {{a_k}{b_k}} \right)} \mathord{\left/
		{\vphantom {{\exp \left( {{a_k}{b_k}} \right)} {\left( {1 + \exp \left( {{a_k}{b_k}} \right)} \right)}}} \right.
		\kern-\nulldelimiterspace} {\left( {1 + \exp \left( {{a_k}{b_k}} \right)} \right)}}$ and ${Y_k} = {{{\xi _k}} \mathord{\left/
		{\vphantom {{{\xi _k}} {\exp \left( {{a_k}{b_k}} \right)}}} \right.
		\kern-\nulldelimiterspace} {\exp \left( {{a_k}{b_k}} \right)}}$ \cite{7934322}. The harvested energy of the $k$-th user is given by
\begin{equation}
	{E_k} = \tau \Xi \left( {{P_k}} \right),\forall k,
\end{equation}
where $\tau $ represents the time duration in the DL phase\footnote{ In order to facilitate analysis, this paper does not consider the user's own circuit power consumption, and only focuses on the power consumption of UL wireless information transmission.}.

Let ${{\bf{\Phi }}_k} = {\rm \mathbb{E}}\left\{ {{{\left( {{{\bf{e}}^H}{{\bf{G}}_k} + {\bf{h}}_{d,k}^H} \right)}^H}\left( {{{\bf{e}}^H}{{\bf{G}}_k} + {\bf{h}}_{d,k}^H} \right)} \right\} \in {\mathbb{C}^{N \times N}}$ be the DL channel covariance matrix of the $k$-th user, then Eq. (17) can be expressed as
\begin{equation}
	{E_k} = \tau \left( {\frac{{{\xi _k}}}{{{X_k}\left( {1 + \exp \left( { - {a_k}\left( {\sum\limits_{l = 1}^L {{\bf{v}}_l^H{{\bf{\Phi }}_k}{{\bf{v}}_l}}  - {b_k}} \right)} \right)} \right)}} - {Y_k}} \right),\forall k.
\end{equation}

In the UL, the $k$-th user uses the energy harvesting in the DL phase to send independent information to HAP, and the transmission signal of the $k$-th user can be expressed as
\begin{equation}
	{x_k} = \sqrt {{p_k}} s_k^{{\rm{ul}}},\forall k,
\end{equation}
where $s_k^{{\rm{ul}}}$ represents the information signal transmitted by the $k$-th user, which is i.i.d CSCG random variable with zero mean and unit variance, i.e., $s_k^{{\rm{ul}}} \sim {\cal C}{\cal N}\left( {0,1} \right),\forall k$. ${p_k}$ is transmit power of the $k$-th user. Considering energy harvesting constraint of the $k$-th user in DL, we have
\begin{equation}
	\left( {1 - \tau } \right){p_k} \le {E_k},\forall k.
\end{equation}
In the UL, through HAP-user channel and HAP-IRS-user channel, the received signal of HAP can be expressed as
\begin{equation}
	\begin{aligned}
		{\bf{y}}& = \sum\limits_{k = 1}^K {\left( {{\bf{h}}_{r,k}^H{\bf{Q}}{{\bf{H}}_{d,r}} + {\bf{h}}_{d,k}^H} \right){x_k}}  + {\bf{n}}\\
		& = \sum\limits_{k = 1}^K {\left( {{\bf{h}}_{r,k}^H{\bf{Q}}{{\bf{H}}_{d,r}} + {\bf{h}}_{d,k}^H} \right)\sqrt {{p_k}} s_k^{{\rm{ul}}} + {\bf{n}}} ,
	\end{aligned}
\end{equation}
where ${\bf{n}} \in {\mathbb{C}^{1 \times N}}$ represents the additive white Gaussian noise (AWGN) introduced by the receiving antenna of the HAP, assuming ${\bf{n}} \sim {\cal C}{\cal N}\left( {{\bf{0}},\sigma _n^2{{\bf{I}}_N}} \right)$. Let ${\bf{q}} = {\left[ {{q_1},...,{q_M}} \right]^T} \in {\mathbb{C}^{M \times 1}}$, then the received signal of HAP can be further given by
\begin{equation}
	{\bf{y}} = \sum\limits_{k = 1}^K {\left( {{{\bf{q}}^H}{{\bf{G}}_k} + {\bf{h}}_{d,k}^H} \right)\sqrt {{p_k}} s_k^{{\rm{ul}}} + {\bf{n}}} .
\end{equation}
In this paper, we assume that a linear receiver is deployed at the HAP to decode the information signal in the UL. Specifically, let ${{\bf{w}}_k} \in {\mathbb{C}^{N \times 1}}$ be the beamforming vector for decoding the information signal $s_k^{{\rm{ul}}}$ transmitted by the $k$-th user. The signal-to-interference-plus-noise ratio (SINR) of the $k$-th user decoded at HAP in UL can be expressed as
\begin{equation}
	{\gamma _k} = \frac{{{p_k}{{\left| {\left( {{{\bf{q}}^H}{{\bf{G}}_k} + {\bf{h}}_{d,k}^H} \right){{\bf{w}}_k}} \right|}^2}}}{{\sum\limits_{i \ne k} {{p_i}{{\left| {\left( {{{\bf{q}}^H}{{\bf{G}}_i} + {\bf{h}}_{d,i}^H} \right){{\bf{w}}_k}} \right|}^2} + \sigma _n^2{{\left\| {{{\bf{w}}_k}} \right\|}^2}} }},\forall k.
\end{equation}
Let ${{\bf{\Psi }}_k} = {\rm \mathbb{E}}\left\{ {{{\left( {{{\bf{q}}^H}{{\bf{G}}_k} + {\bf{h}}_{d,k}^H} \right)}^H}\left( {{{\bf{q}}^H}{{\bf{G}}_k} + {\bf{h}}_{d,k}^H} \right)} \right\} \in {\mathbb{C}^{N \times N}}$ be the UL channel covariance matrix of the $k$-th user\footnote{$\bf{\Phi_k}$ and $\bf{\Psi_k}$ represent the DL and UL channel covariance matrices, respectively. The definition of the channel covariance matrix is to take the expectation of the product of the channel gain and its conjugate transpose. However, in calculation process, due to the assumption that the channel is constant in each transmission time duration $T$, we can ignore the operation of the expectation operator.}, then Eq. (23) can be expressed as
\begin{equation}
	{\gamma _k} = \frac{{{p_k}{\bf{w}}_k^H{{\bf{\Psi }}_k}{{\bf{w}}_k}}}{{\sum\limits_{i \ne k} {{p_i}{\bf{w}}_k^H{{\bf{\Psi }}_i}{{\bf{w}}_k} + \sigma _n^2{\bf{w}}_k^H{{\bf{w}}_k}} }},\forall k.
\end{equation}

In this paper, we consider the setup where perfect CSI of DL and UL cannot be obtained. Specifically, for DL transmission, it is assumed that the channel covariance matrix is ${{\bf{\Phi }}_k} + \Delta {\bf{E}}_k^{{\rm{DL}}}$, where ${{\bf{\Phi }}_k} \in {\mathbb{C}^{N \times N}}$ is the estimated channel covariance matrix in the DL, and $\Delta {\bf{E}}_k^{{\rm{DL}}} \in {\mathbb{C}^{N \times N}}$ denotes covariance uncertainty matrix of the corresponding HAP-user channel and HAP-IRS-user channel in the DL, which describes the difference between the true value and the estimated value of two channels \cite{4350296,chalise2004robust}. Similarly, for UL transmission, it is assumed that the channel covariance matrix is ${{\bf{\Psi }}_k} + \Delta {\bf{E}}_k^{{\rm{UL}}}$, where ${{\bf{\Psi }}_k} \in {\mathbb{C}^{N \times N}}$ is the estimated channel covariance matrix in the UL and $\Delta {\bf{E}}_k^{{\rm{UL}}} \in {\mathbb{C}^{N \times N}}$ represents covariance uncertainty matrix of the corresponding HAP-user channel and HAP-IRS-user channel in the UL. We assume that the statistical information of $\Delta {\bf{E}}_k^{{\rm{DL}}}$ and $\Delta {\bf{E}}_k^{{\rm{UL}}}$ cannot be obtained, but certain threshold can be obtained, i.e.,
\begin{equation}
	\left\| {\Delta {\bf{E}}_k^{{\rm{DL}}}} \right\| \le \varepsilon _k^{{\rm{DL}}},\forall k,
\end{equation}
and
\begin{equation}
	\left\| {\Delta {\bf{E}}_k^{{\rm{UL}}}} \right\| \le \varepsilon _k^{{\rm{UL}}},\forall k,
\end{equation}
where $\left\|  \cdot  \right\|$ represents the matrix norm. It is worth noting that ${{\bf{\Phi }}_k}$, ${{\bf{\Psi }}_k}$, ${\Delta {\bf{E}}_k^{{\rm{DL}}}}$ and ${\Delta {\bf{E}}_k^{{\rm{UL}}}}$ are all Hermitian matrices, Eq. (18) and Eq. (24) can be respectively expressed as
\begin{equation}
	{E_k} = \tau \left( {\frac{{{\xi _k}}}{{{X_k}\left( {1 + \exp \left( { - {a_k}\left( {\sum\limits_{l = 1}^L {{\bf{v}}_l^H\left( {{{\bf{\Phi }}_k} + \Delta {\bf{E}}_k^{{\rm{DL}}}} \right){{\bf{v}}_l}}  - {b_k}} \right)} \right)} \right)}} - {Y_k}} \right),\forall k,
\end{equation}
and
\begin{equation}
	{\gamma _k} = \frac{{{p_k}{\bf{w}}_k^H\left( {{{\bf{\Psi }}_k} + \Delta {\bf{E}}_k^{{\rm{UL}}}} \right){{\bf{w}}_k}}}{{\sum\limits_{i \ne k} {{p_i}{\bf{w}}_k^H\left( {{{\bf{\Psi }}_i} + \Delta {\bf{E}}_i^{{\rm{UL}}}} \right){{\bf{w}}_k} + \sigma _n^2{\bf{w}}_k^H{{\bf{w}}_k}} }},\forall k.
\end{equation}
\subsection{Problem formulation}
In this paper, we define ${{\bf{V}}^{{\rm{dl}}}} = \left[ {{{\bf{v}}_1},...,{{\bf{v}}_L}} \right]$, ${{\bf{W}}^{{\rm{ul}}}} = \left[ {{{\bf{w}}_1},...,{{\bf{w}}_K}} \right]$ and ${\bf{P}} = \left[ {{p_1},...,{p_K}} \right]$. We jointly optimize the time allocation $\tau $, the HAP transmit energy beamforming ${{\bf{V}}^{{\rm{dl}}}}$ in the DL, IRS energy reflection coefficient ${\bf{e}}$, the HAP receiving beamforming ${{\bf{W}}^{{\rm{ul}}}}$ in the UL, IRS information reflection coefficient $\bf{q}$, and user transmit power allocation $\bf{P}$ in the UL to minimize the transmit energy of the HAP. The optimization problem can be expressed as follows,
\begin{subequations}
	\begin{align}
		\left( {{\textrm{P1}}} \right){\rm{~~~~~}}&\mathop {\min }\limits_{\tau ,{{\bf{V}}^{{\rm{dl}}}},{{\bf{W}}^{{\rm{ul}}}},{\bf{P}},{\bf{e}},{\bf{q}}} {\rm{ }}\tau \sum\limits_{l = 1}^L {{{\left\| {{{\bf{v}}_l}} \right\|}^2}} , \\
		\rm{s.t.}\qquad &0 \le \tau  \le 1,\\
		&\sum\limits_{l = 1}^L {{{\left\| {{{\bf{v}}_l}} \right\|}^2}}  \le {P_{\max }},\\
		&\left( {1 - \tau } \right){p_k} \le {E_k},\forall \left\| {\Delta {\bf{E}}_k^{{\rm{DL}}}} \right\| \le \varepsilon _k^{{\rm{DL}}},\forall k,\\
		&{\gamma _k} \ge {\gamma _{th}},\forall \left\| {\Delta {\bf{E}}_k^{{\rm{UL}}}} \right\| \le \varepsilon _k^{{\rm{UL}}},\forall k,\\
		&{\left| {{e_m}} \right|^2} = 1,\forall m,\\
		&{\left| {{q_m}} \right|^2} = 1,\forall m.
	\end{align}
\end{subequations}
where (29b) is the time allocation constraint, and (29c) is the maximum transmit power constraint at the HAP in the DL phase, and (29d) means when any covariance uncertainty matrix satisfies $\left\| {\Delta {\bf{E}}_k^{{\rm{DL}}}} \right\| \le \varepsilon _k^{{\rm{DL}}}$, the energy constraint of the $k$-th user in the DL holds, and (29e) means when any covariance uncertainty matrix satisfies the $\left\| {\Delta {\bf{E}}_k^{{\rm{UL}}}} \right\| \le \varepsilon _k^{{\rm{UL}}}$, the user's SINR constraint should be satisfied in the UL, and (29f) and (29g) are the unit modulus 1 constraint of the reflection coefficient of the IRS in the DL and UL, respectively.

\section{Robust beamforming design and time allocation algorithm for IRS-assisted WPCN}
\subsection{Problem Transformation}
Apparently, the optimization problem (P1) is non-convex and cannot be solved directly. First, transform constraints (29d) and (29e) according to $\textbf{Lemma~1}$.

$\textbf{Lemma~1:}$ \emph{For any Hermitian matrix ${\bf{A}}$ and ${\bf{B}}$, if ${\bf{B}}$ satisfies $\left\| {\bf{B}} \right\| \le \varepsilon $, then 
\begin{equation}
	\mathop {\max }\limits_{\left\| {\bf{B}} \right\| \le \varepsilon } {\rm{tr}}\left( {{\bf{AB}}} \right) = \varepsilon {\left\| {\bf{A}} \right\|_ * }
\end{equation}
holds, where ${\left\| {\bf{A}} \right\|_ * }$ denotes the dual norm of matrix ${\bf{A}}$.}

\emph{Proof:} According to the definition of the dual norm, for a complex matrix ${\bf{X}}$ whose norm is less than 1, the maximum value of the inner product of the complex matrix ${\bf{Z}}$ and ${\bf{X}}$ is the dual norm of ${\bf{Z}}$, i.e.,
\begin{equation}
	{\left\| {\bf{Z}} \right\|_ * } = \sup \left\{ {{\rm{tr}}\left( {{{\bf{Z}}^H}{\bf{X}}} \right)\left| {\left\| {\bf{X}} \right\| \le 1} \right.} \right\}.
\end{equation}
From the definition of the dual norm, it can be obtained that for complex matrix ${\bf{X}}$ and ${\bf{Z}}$, the following inequalities hold
\begin{equation}
	{\rm{tr}}\left( {{{\bf{Z}}^H}{\bf{X}}} \right) \le \left\| {\bf{X}} \right\|{\left\| {\bf{Z}} \right\|_ * }.
\end{equation}
Since both matrices ${\bf{A}}$ and ${\bf{B}}$ are Hermitian matrices, there are 
\begin{equation}
	{\rm{tr}}\left( {{\bf{AB}}} \right) \le \left\| {\bf{B}} \right\|{\left\| {\bf{A}} \right\|_ * } \le \varepsilon {\left\| {\bf{A}} \right\|_ * }.
\end{equation}
Thus
\begin{equation}
	\mathop {\max }\limits_{\left\| {\bf{B}} \right\| \le \varepsilon } {\rm{tr}}\left( {{\bf{AB}}} \right) = \varepsilon {\left\| {\bf{A}} \right\|_ * }
\end{equation}
holds. The proof of $\textbf{Lemma~1}$ is completed. $\hfill\blacksquare$

If we use the spectral norm of ${\bf{B}}$, i.e., the maximum singular value of ${\bf{B}}$ satisfies ${\sigma _{\max }}\left( {\bf{B}} \right) \le \varepsilon $. Since the dual norm of the spectral norm is the nuclear norm, which is equal to the sum of singular values, then the dual norm of ${\bf{A}}$ can be expressed as ${\left\| {\bf{A}} \right\|_ * } = \sum {\sigma \left( {\bf{A}} \right)} $.

Further, according to ${\bf{v}}_l^H\Delta {\bf{E}}_k^{{\rm{DL}}}{{\bf{v}}_l} = {\rm{tr}}\left( {{\bf{v}}_l^H\Delta {\bf{E}}_k^{{\rm{DL}}}{{\bf{v}}_l}} \right) = {\rm{tr}}\left( {\Delta {\bf{E}}_k^{{\rm{DL}}}{{\bf{v}}_l}{\bf{v}}_l^H} \right) = {\rm{tr}}\left( {\Delta {\bf{E}}_k^{{\rm{DL}}}{{\bf{V}}_l}} \right)$, where ${{\bf{V}}_l} = {{\bf{v}}_l}{\bf{v}}_l^H \succeq 0$ and ${\rm{rank}}\left( {{{\bf{V}}_l}} \right) = 1$. We use the spectral norm of ${\Delta {\bf{E}}_k^{{\rm{UL}}}}$ for $\left\| {\Delta {\bf{E}}_k^{{\rm{DL}}}} \right\| \le \varepsilon _k^{{\rm{DL}}}$. According to the $\textbf{Lemma~1}$, we have 
\begin{equation}
	\mathop {\max }\limits_{\left\| {\Delta {\bf{E}}_k^{{\rm{DL}}}} \right\| \le {\varepsilon _k}} {\rm{tr}}\left( {\Delta {\bf{E}}_k^{{\rm{DL}}}{{\bf{V}}_l}} \right) = \varepsilon _k^{{\rm{DL}}}{\left\| {{{\bf{V}}_l}} \right\|_*} = \varepsilon _k^{{\rm{DL}}}{\rm{tr}}\left( {{{\bf{V}}_l}} \right).
\end{equation}
In the worst case, Eq. (27) can be denoted by
\begin{equation}
	{E_k} = \tau \left( {\frac{{{\xi _k}}}{{{X_k}\left( {1 + \exp \left( { - {a_k}\left( {\sum\limits_{l = 1}^L {{\rm{tr}}\left( {\left( {{{\bf{\Phi }}_k} - \varepsilon _k^{{\rm{DL}}}{{\bf{I}}_N}} \right){{\bf{V}}_l}} \right)}  - {b_k}} \right)} \right)} \right)}} - {Y_k}} \right),\forall k.
\end{equation}
Similarly, we have ${\bf{w}}_k^H\Delta {\bf{E}}_k^{{\rm{UL}}}{{\bf{w}}_k} = {\rm{tr}}\left( {{\bf{w}}_k^H\Delta {\bf{E}}_k^{{\rm{UL}}}{{\bf{w}}_k}} \right) = {\rm{tr}}\left( {\Delta {\bf{E}}_k^{{\rm{UL}}}{{\bf{w}}_k}{\bf{w}}_k^H} \right) = {\rm{tr}}\left( {\Delta {\bf{E}}_k^{{\rm{UL}}}{{\bf{W}}_k}} \right)$, where ${{\bf{W}}_k} = {{\bf{w}}_k}{\bf{w}}_k^H \succeq 0$ and ${\rm{rank}}\left( {{{\bf{W}}_k}} \right) = 1$. Thus,
\begin{equation}
	\mathop {\max }\limits_{\left\| {\Delta {\bf{E}}_k^{{\rm{UL}}}} \right\| \le {\varepsilon _k}} {\rm{tr}}\left( {\Delta {\bf{E}}_k^{{\rm{UL}}}{{\bf{W}}_k}} \right) = \varepsilon _k^{{\rm{UL}}}{\left\| {{{\bf{W}}_k}} \right\|_*} = \varepsilon _k^{{\rm{UL}}}{\rm{tr}}\left( {{{\bf{W}}_k}} \right).
\end{equation}
In the worst case, Eq. (28) can also be expressed as
\begin{equation}
	{\gamma _k} = \frac{{{p_k}{\rm{tr}}\left( {\left( {{{\bf{\Psi }}_k} - \varepsilon _k^{{\rm{UL}}}{{\bf{I}}_N}} \right){{\bf{W}}_k}} \right)}}{{\sum\limits_{i \ne k} {{p_i}{\rm{tr}}\left( {\left( {{{\bf{\Psi }}_i} + \varepsilon _i^{{\rm{UL}}}{{\bf{I}}_N}} \right){{\bf{W}}_k}} \right) + \sigma _n^2{\rm{tr}}\left( {{{\bf{W}}_k}} \right)} }},\forall k.
\end{equation}
Let $\left\{ {{{\bf{V}}_l}} \right\}$ and $\left\{ {{{\bf{W}}_k}} \right\}$ denote the set of ${{{\bf{V}}_l}}$ and ${{{\bf{W}}_k}}$, respectively. Therefore, the problem (P1) can be transformed into the problem (P2) as follows
\begin{subequations}
	\begin{align}
		\left( {{\textrm{P2}}} \right){\rm{~~~~~}}&\mathop {\min }\limits_{\tau ,\left\{ {{{\bf{V}}_l}} \right\},\left\{ {{{\bf{W}}_k}} \right\},{\bf{P}},{\bf{e}},{\bf{q}}} {\rm{ }}\tau \sum\limits_{l = 1}^L {{\rm{tr}}\left( {{{\bf{V}}_l}} \right)} , \\
		\rm{s.t.}\qquad &0 \le \tau  \le 1,\\
		&\sum\limits_{l = 1}^L {{\rm{tr}}\left( {{{\bf{V}}_l}} \right)}  \le {P_{\max }},\\
		&\left( {1 - \tau } \right){p_k} \le {E_k},\forall k,\\
		&{\gamma _k} \ge {\gamma _{th}},\forall k,\\
		&{\left| {{e_m}} \right|^2} = 1,\forall m,\\
		&{\left| {{q_m}} \right|^2} = 1,\forall m,\\
		&{{\bf{V}}_l} \succeq  0,\forall l,{\rm{rank}}\left( {{{\bf{V}}_l}} \right) = 1,\forall l,\\
		&{{\bf{W}}_k}\succeq  0,\forall k,{\rm{rank}}\left( {{{\bf{W}}_k}} \right) = 1,\forall k.
	\end{align}
\end{subequations}
\subsection{Problem Solution}
In this sub-section, we apply AO technique to divide the problem (P2) into three sub-problems to obtain its solution. Firstly, given the time allocation and IRS reflection coefficient, a robust beamforming design of HAP and user transmit power allocation algorithm is proposed. Next, when time allocation and HAP beamforming design are fixed, a robust beamforming design of IRS is given. Finally, based on the HAP beamforming design and the IRS beamforming design already obtained, the time allocation can be easily obtained.
\subsubsection{Robust beamforming design of HAP and user transmit power allocation}
First of all, let ${\bf{S}} = \sum\limits_{l = 1}^L {{{\bf{V}}_l}} $. For the sub-problem 1, given time allocation $\tau $ and the IRS beamforming design ${\bf{e}}$, ${\bf{q}}$, the problem (P2) can be transformed into the problem (P3) as follows
\begin{subequations}
	\begin{align}
		\left( {{\textrm{P3}}} \right){\rm{~~~~}}&\mathop {\min }\limits_{{\bf{S}},\left\{ {{{\bf{W}}_k}} \right\},{\bf{P}}} {\rm{ }}\tau {\rm{tr}}\left( {\bf{S}} \right), \\
		\rm{s.t.}\qquad &{\rm{tr}}\left( {\bf{S}} \right) \le {P_{\max }},\\
		&\left( {1 - \tau } \right){p_k} \le \tau \left( {\frac{{{\xi _k}}}{{{X_k}\left( {1 + \exp \left( { - {a_k}\left( {{\rm{tr}}\left( {\left( {{{\bf{\Phi }}_k} - \varepsilon _k^{{\rm{DL}}}{{\bf{I}}_N}} \right){\bf{S}}} \right) - {b_k}} \right)} \right)} \right)}} - {Y_k}} \right),\forall k,\\
		&{\gamma _k} \ge {\gamma _{th}},\forall k,\\
		&{{\bf{W}}_k} \succeq 0,\forall k,\\
		&{\rm{rank}}\left( {{{\bf{W}}_k}} \right) = 1,\forall k,\\
		&{\bf{S}} \succeq 0.
	\end{align}
\end{subequations}
The optimization problem (P3) is still non-convex. Below we divide it into three parts to solve. Fixed $\left\{ {{{\bf{W}}_k}} \right\}$ and ${\bf{P}}$, the problem (P3) is transformed into the problem into (P3.1), which can be given by
\begin{subequations}
	\begin{align}
		\left( {{\textrm{P3.1}}} \right){\rm{~~~~}}&\mathop {\min }\limits_{\bf{S}} {\rm{ ~~}}\tau {\rm{tr}}\left( {\bf{S}} \right), \\
		\rm{s.t.}\qquad & \textrm {(40b), (40g)},\\
		& {\Xi ^{ - 1}}\left( {\frac{{1 - \tau }}{\tau }{p_k}} \right) \le {\rm{tr}}\left( {\left( {{{\bf{\Phi }}_k} - \varepsilon _k^{{\rm{DL}}}{{\bf{I}}_N}} \right){\bf{S}}} \right),\forall k,
	\end{align}
\end{subequations}
where ${\Xi ^{ - 1}}\left( x \right)$ represents the inverse function of $\Xi \left( x \right)$, ${\Xi ^{ - 1}}\left( x \right) = {b_k} - \frac{{\ln \left( {{{{\xi _k}} \mathord{\left/
					{\vphantom {{{\xi _k}} {\left( {\left( {x + {Y_k}} \right){X_k}} \right)}}} \right.
					\kern-\nulldelimiterspace} {\left( {\left( {x + {Y_k}} \right){X_k}} \right)}} - 1} \right)}}{{{a_k}}}$. It can be seen that the problem is a semi-definite programming (SDP) problem and can be solved by applying the CVX toolbox \cite{grant2014cvx}. Let ${{\bf{S}}^ * }$ represent the optimal solution of the problem (P3.1), then the optimal $L = {\rm{rank}}\left( {{{\bf{S}}^ * }} \right)$ is the number of energy beams in DL. In addition, $\left\{ {{\bf{v}}_1^ * ,...,{\bf{v}}_L^ * } \right\}$ can be obtained by eigenvalue decomposition of ${{\bf{S}}^ * }$.

$\textbf{Lemma~2:}$ \emph{Assuming that the separable SDP problem $({\cal P}){\rm{ }}$ and its dual problem $({\cal D}){\rm{ }}$ are both solvable, the problem $({\cal P}){\rm{ }}$ always has an optimal solution ${\bf{X}}_a^ * $ that satisfies the following conditions
\begin{equation}
	\sum\limits_{a = 1}^A {{{\left( {{\rm{rank}}\left( {{\bf{X}}_a^ * } \right)} \right)}^2} \le B} ,
\end{equation}
where $A$ represents the number of optimization variables, and $B$ represents the number of constraints \cite{5233822}.}

Further, given ${\bf{S}}$ and ${\bf{P}}$, the problem (P3) can be converted to a feasibility-check problem (P3.2) as follows
\begin{subequations}
	\begin{align}
		\left( {{\textrm{P3.2}}} \right){\rm{~~~~}}&{\rm{find~~}}{{\bf{W}}_k}, \\
		\rm{s.t.}\qquad & \textrm {(40d), (40e), (40f)}.
	\end{align}
\end{subequations}
We first apply SDR to relax the non-convex rank-one constraint (40f). Hence, the problem (P3.2) can be converted to the problem (P3.2.1), which can be represented as
\begin{subequations}
	\begin{align}
		\left( {{\textrm{P3.2.1}}} \right){\rm{~~~~}}&{\rm{find~~}}{{\bf{W}}_k}, \\
		\rm{s.t.}\qquad & \textrm {(40d), (40e)}.
	\end{align}
\end{subequations}
We can see that the problem (P3.2.1) is an SDP problem, which can be solved by applying the CVX toolbox \cite{grant2014cvx}. According to $\textbf{Lemma~2}$, the problem (P3.2.1) and its dual problem can be solved, the number of optimization variables and the number of constraints are both $K$, so the problem (P3.2.1) must have a rank-one solution. Herein, the optimal ${\bf{w}}_k^ * $ can be obtained by eigenvalue decomposition of ${\bf{W}}_k^ * $.

Finally, when ${\bf{S}}$ and $\left\{ {{{\bf{W}}_k}} \right\}$ are fixed, the problem (P3) is also converted to a feasibility-check problem (P3.3) as follows
\begin{subequations}
	\begin{align}
		\left( {{\textrm{P3.3}}} \right){\rm{~~~~}}&{\rm{find~~ }}{p_k}, \\
		\rm{s.t.}\qquad & \textrm {(40c), (40d)}.
	\end{align}
\end{subequations}
This problem is an LP problem, which can be solved by applying the CVX toolbox \cite{grant2014cvx}.
Therefore, the robust beamforming design of HAP and user transmit power allocation algorithm can be shown as $\textbf{Algorithm 1}$.
\begin{algorithm}[H]
	\caption{The Robust Beamforming Design of HAP and User Transmit Power Allocation Algorithm} 
	\begin{algorithmic}[1]
		\State $\textbf{Input:}$ $\tau^0 $, ${\bf{e}}^0$, ${\bf{q}}^0$, convergence threshold $\epsilon$ and iteration index $t = 0$.
		\Repeat 
		\State For the $t$-th ineration, given ${{\bf{W}}_k}$ and $p_k$, solve the problem (P3.1) to obtain ${{\bf{S}}^ * }$, and $\left\{ {{\bf{v}}_1^ * ,...,{\bf{v}}_L^ * } \right\}$ can be obtained by eigenvalue decomposition of ${{\bf{S}}^ * }$.
		\State For the $t$-th ineration, given ${\bf{S}}$ and $p_k$, solve the feasibility-check problem (P3.2.1), ${\bf{w}}_k^ * $ can be obtained by eigenvalue decomposition of ${\bf{W}}_k^ * $.
		\State For the $t$-th ineration, given ${\bf{S}}$ and ${{\bf{W}}_k}$, solve the feasibility-check problem (P3.3) to obtain ${p}_k^ * $.
		\State Update $t = t + 1$.
		\Until the optimization objective function value meets the convergence threshold $\epsilon$.
		\State $\textbf{Output:}$ HAP energy beamforming vector ${\bf{v}}_l^ * $, receiving beamforming vector ${\bf{w}}_k^ * $ and user transmit power ${p}_k^ * $.
	\end{algorithmic}
\end{algorithm}
\subsubsection{Robust beamforming design of IRS}
For the sub-problem 2, when the time allocation $\tau $ and HAP beamforming design $\left\{ {{{\bf{V}}_l}} \right\}$, $\left\{ {{{\bf{W}}_k}} \right\}$, ${{\bf{P}}}$ are given, the problem (P2) can be converted to the feasibility-check problem (P4), which can be represented as
\begin{subequations}
	\begin{align}
		\left( {{\textrm{P4}}} \right){\rm{~~~~}}&{\rm{find ~~ }}{\bf{e}}{\rm{,}}{\bf{q}}, \\
		\rm{s.t.}\qquad &{\Xi ^{ - 1}}\left( {\frac{{1 - \tau }}{\tau }{p_k}} \right) \le \sum\limits_{l = 1}^L {{\rm{tr}}\left( {\left( {{{\bf{\Phi }}_k} - {\varepsilon ^{\rm DL}_k}{{\bf{I}}_N}} \right){{\bf{V}}_l}} \right)} ,\forall k,\\
		&{p_k}\left( {{\rm{tr}}\left( {{{\bf{\Psi }}_k}{{\bf{W}}_k}} \right)\! \!-\! {\varepsilon ^{\rm UL}_k}{\rm{tr}}\left( {{{\bf{I}}_N}{{\bf{W}}_k}} \right)} \right){\rm{ }} \!\!-\! {\gamma _{th}}\sum\limits_{i \ne k} {{p_i}\left( {{\rm{tr}}\left( {{{\bf{\Psi }}_i}{{\bf{W}}_k}} \right)\!\! -\! {\varepsilon ^{\rm UL}_i}{\rm{tr}}\left( {{{\bf{I}}_N}{{\bf{W}}_k}} \right)} \right)} \! \!\ge\! {\gamma _{th}}\sigma _n^2{\rm{tr}}\left( {{{\bf{W}}_k}} \right),\!\forall k\\
		&{\left| {{e_m}} \right|^2} = 1,\forall m,\\
        &{\left| {{q_m}} \right|^2} = 1,\forall m.
    \end{align}
\end{subequations}
Due to the constraint (46d) and (46e) of the IRS phase, the problem (P4) is obviously non-convex. Therefore, we divide it into two parts to solve. Fixed ${\bf{q}}$, the problem (P4) can be converted to the problem (P4.1) as follows
\begin{subequations}
	\begin{align}
		\left( {{\textrm{P4.1}}} \right){\rm{~~~~}}&{\rm{find~~ }}{\bf{e}}, \\
		\rm{s.t.}\qquad & \sum\limits_{l = 1}^L {{\rm{tr}}\left( {{{\bf{\Phi }}_k}{{\bf{V}}_l}} \right)}  \ge {\Xi ^{ - 1}}\left( {\frac{{1 - \tau }}{\tau }{p_k}} \right) + {\varepsilon ^{\rm DL}_k}\sum\limits_{l = 1}^L {{\rm{tr}}\left( {{{\bf{I}}_N}{{\bf{V}}_l}} \right)} ,\forall k,\\
		&{\left| {{e_m}} \right|^2} = 1,\forall m.
	\end{align}
\end{subequations}
The left side of constraint (46b) $\sum\limits_{l = 1}^L {{\rm{tr}}\left( {{{\bf{\Phi }}_k}{{\bf{V}}_l}} \right)}  = {\sum\limits_{l = 1}^L {\left| {\left( {{{\bf{e}}^H}{{\bf{G}}_k} + {\bf{h}}_{d,k}^H} \right){{\bf{v}}_l}} \right|} ^2} = {\sum\limits_{l = 1}^L {\left| {{{\bf{e}}^H}{{\bf{a}}_{kl}} + {b_{kl}}} \right|} ^2}$, where ${{\bf{a}}_{kl}} = {{\bf{G}}_k}{{\bf{v}}_l} \in {\mathbb{C}^{M \times 1}}$ and ${b_{kl}} = {\bf{h}}_{d,k}^H{{\bf{v}}_l} \in {\mathbb{C}} $. Introduce auxiliary matrices as follows
\begin{equation}
	{{\bf{R}}_{kl}} = \left[ {\begin{array}{*{20}{c}}
			{{{\bf{a}}_{kl}}{\bf{a}}_{kl}^H}&{{{\bf{a}}_{kl}}b_{kl}^H}\\
			{{\bf{a}}_{kl}^H{b_{kl}}}&0
	\end{array}} \right],{\bf{\bar e}} = \left[ {\begin{array}{*{20}{c}}
			{\bf{e}}\\
			1
	\end{array}} \right].
\end{equation}
Thus 
\begin{equation}
	\begin{aligned}
		&{\sum\limits_{l = 1}^L {\left| {{{\bf{e}}^H}{{\bf{a}}_{kl}} + {b_{kl}}} \right|} ^2} = \sum\limits_{l = 1}^L {\left( {{{{\bf{\bar e}}}^H}{{\bf{R}}_{kl}}{\bf{\bar e}} + {{\left| {{b_{kl}}} \right|}^2}} \right)} \\
		& = \sum\limits_{l = 1}^L {\left( {{\rm{tr}}\left( {{{\bf{R}}_{kl}}{\bf{\bar E}}} \right) + {{\left| {{b_{kl}}} \right|}^2}} \right)},\\
	\end{aligned}
\end{equation}
where ${\bf{\bar E}} = {\bf{\bar e}}{{\bf{\bar e}}^H}\succeq 0$ and ${\rm{rank}}\left( {{\bf{\bar E}}} \right) = 1$. Then the problem (P4.1) can be rewritten as the problem (P4.1.1), which can be given by
\begin{subequations}
	\begin{align}
		\left( {{\textrm{P4.1.1}}} \right){\rm{~~~~}}&{\rm{find~~ }}{\bf{\bar E}}, \\
		\rm{s.t.}\qquad & \sum\limits_{l = 1}^L {{\rm{tr}}\left( {{{\bf{R}}_{kl}}{\bf{\bar E}}} \right)}  \ge {\Xi ^{ - 1}}\left( {\frac{{1 - \tau }}{\tau }{p_k}} \right) + {\varepsilon ^{\rm DL}_k}\sum\limits_{l = 1}^L {{\rm{tr}}\left( {{{\bf{I}}_N}{{\bf{V}}_l}} \right) - \sum\limits_{l = 1}^L {{{\left| {{b_{kl}}} \right|}^2}} } ,\forall k,\\
		&{\bf{\bar E}}\succeq0,\\
		&{\rm{rank}}\left( {{\bf{\bar E}}} \right) = 1,\\
		&{{{\bf{\bar E}}}_{m,m}} = 1,m = 1,...,M + 1.
	\end{align}
\end{subequations}
As for the problem (P4.1.1), we usually use SDR to relax the non-convex rank-one constraint (49d), and the problem is converted to an SDP problem, which can be solved by using the CVX toolbox \cite{grant2014cvx}. Let ${{\bf{\bar E}}^ * }$ denote the optimal solution of the SDP problem. If the obtained ${{\bf{\bar E}}^ * }$ is rank-one, then ${{\bf{\bar e}}^ * }$ can be obtained by eigenvalue decomposition of ${{\bf{\bar E}}^ * }$. Conversely, since the number of optimization variables for this problem is one, and the number of constraints is $M+K$, it may not be guaranteed to obtain a rank-one solution according to the $\textbf{Lemma~2}$. Gaussian randomization is usually used to obtain a sub-optimal solution to the original problem. However, due to the large number of reflection elements in the IRS, i.e., the scale of the optimization problem is larger, it is difficult to use Gaussian randomization to return a rank-one solution. In this paper, we apply the DC programming to convert the non-convex rank-one constraint and obtain the solution of the problem (P4.1.1) \cite{9352968}.

$\textbf{Proposition 1:}$ \emph{For the positive semi-definite matrix ${\bf{M}} \in {\mathbb{C}^{N \times N}}$, ${\rm{tr}}\left( {\bf{M}} \right) > 0$, the rank-one constraint can be equivalent to the difference between two convex functions, which can be given by
\begin{equation}
	{\rm{rank}}\left( {\bf{M}} \right) = 1 \Leftrightarrow {\rm{tr}}\left( {\bf{M}} \right) - {\left\| {\bf{M}} \right\|_2}=0,
\end{equation}
where ${\rm{tr}}\left( {\bf{M}} \right) = \sum\limits_{n = 1}^N {{\sigma _n}\left( {\bf{M}} \right)} $, ${\left\| {\bf{M}} \right\|_2} = {\sigma _1}\left( {\bf{M}} \right)$ is spectral norm, and ${\sigma _n}\left( {\bf{M}} \right)$ represents the $n$-th largest singular value of matrix ${\bf{M}}$. }

According to $\textbf{Proposition 1}$, the problem (P4.1.1) can be rewritten as the problem (P4.1.2) as follows
\begin{subequations}
	\begin{align}
		\left( {{\textrm{P4.1.2}}} \right){\rm{~~~~}}&\min {\rm{~tr}}\left( {{\bf{\bar E}}} \right) - {\left\| {{\bf{\bar E}}} \right\|_2}, \\
		\rm{s.t.}\qquad &  \textrm {(50b), (50c), (50e)}.
	\end{align}
\end{subequations}
The problem (P4.1.2) is still a non-convex problem because $ - {\left\| {{\bf{\bar E}}} \right\|_2}$ is concave. The core of the DC programming is to convert the problem to a convex optimization problem by linearizing the concave term. Specifically, we need to solve the following problem in the $t$-th iteration,
\begin{subequations}
	\begin{align}
		\left( {{\textrm{P4.1.3}}} \right){\rm{~~~~}}&\min {\rm{ ~ tr}}\left( {{\bf{\bar E}}} \right) - \left\langle {\partial {{\left\| {{{{\bf{\bar E}}}^{t - 1}}} \right\|}_2},{\bf{\bar E}}} \right\rangle , \\
		\rm{s.t.}\qquad & \textrm {(49b), (49c), (49e)},
	\end{align}
\end{subequations}
where ${{\bf{\bar E}}^{t - 1}}$ is the solution obtained at the $t-1$ iteration, and $\partial {\left\| {{{{\bf{\bar E}}}^{t - 1}}} \right\|_2}$ denotes the subgradient of the spectral norm at the $t-1$ iteration, which can be calculated by $\textbf{Proposition 2}$. The problem (P4.1.3) is a convex optimization problem, which can be solved by applying the CVX toolbox \cite{grant2014cvx}. By solving problem (P4.1.3) iteratively until the optimal value is zero, we can obtain a rank-one solution. In numerical simulation, we usually set the stopping criterion ${\rm{tr}}\left( {{\bf{\bar E}}} \right) - {\left\| {{\bf{\bar E}}} \right\|_2} \le \epsilon $, where $\epsilon $ is a sufficiently small constant. The convergence of the DC programming can be guaranteed.

$\textbf{Proposition 2:}$ \emph{For the positive semi-definite matrix ${\bf{M}}$, the subgradient $\partial {\left\| {\bf{M}} \right\|_2}$ of the matrix spectral norm can be obtained by ${{\bf{m}}_1}{\bf{m}}_1^H$, where ${{\bf{m}}_1} \in {\mathbb{C}^{{\rm{1}} \times N}}$ is the eigenvector corresponding to the largest singular value of the matrix ${\bf{M}}$ \cite{tao1997convex}.}

Similarly, if ${\bf{e}}$ is fixed, the problem (P4) can also be converted to the problem (P4.2) as follows
\begin{subequations}
	\begin{align}
		\left( {{\textrm{P4.2}}} \right){\rm{~~~~}}&{\rm{find~~ }}{\bf{q}}, \\
		\rm{s.t.}\qquad & \begin{array}{l}
			{p_k}{\rm{tr}}\left( {{{\bf{\Psi }}_k}{{\bf{W}}_k}} \right) - {\gamma _{th}}\sum\limits_{i \ne k} {{p_i}{\rm{tr}}\left( {{{\bf{\Psi }}_i}{{\bf{W}}_k}} \right)}  \ge {\gamma _{th}}\sigma _n^2{\rm{tr}}\left( {{{\bf{W}}_k}} \right) + {p_k}{\varepsilon ^{\rm UL}_k}{\rm{tr}}\left( {{{\bf{I}}_N}{{\bf{W}}_k}} \right)\\
			- {\gamma _{th}}\sum\limits_{i \ne k} {{p_i}{\varepsilon  ^{\rm UL}_i}{\rm{tr}}\left( {{{\bf{I}}_N}{{\bf{W}}_k}} \right)} ,\forall k,
		\end{array}\\
		&{\left| {{q_m}} \right|^2} = 1,\forall m.
	\end{align}
\end{subequations}
The left side of constraint (53b) ${\rm{tr}}\left( {{{\bf{\Psi }}_i}{{\bf{W}}_k}} \right) = {\left| {\left( {{{\bf{q}}^H}{{\bf{G}}_i} + {\bf{h}}_{d,i}^H} \right){{\bf{w}}_k}} \right|^2} = {\left| {{{\bf{q}}^H}{{\bf{c}}_{ik}} + {d_{ik}}} \right|^2}$, where ${{\bf{c}}_{ik}} = {{\bf{G}}_i}{{\bf{w}}_k} \in {\mathbb{C}^{M \times 1}}$ and ${d_{ik}} = {\bf{h}}_{d,i}^H{{\bf{w}}_k} \in \mathbb{C}$. Introduce auxiliary matrices as follows
\begin{equation}
	{{\bf{T}}_{ik}} = \left[ {\begin{array}{*{20}{c}}
			{{{\bf{c}}_{ik}}{\bf{c}}_{ik}^H}&{{{\bf{c}}_{ik}}d_{ik}^H}\\
			{{\bf{c}}_{ik}^H{d_{ik}}}&0
	\end{array}} \right],{\bf{\bar q}} = \left[ {\begin{array}{*{20}{c}}
			{\bf{q}}\\
			1
	\end{array}} \right].
\end{equation}
Thus 
\begin{equation}
	{\left| {{{\bf{q}}^H}{{\bf{c}}_{ik}} + {d_{ik}}} \right|^2} = {{\bf{\bar q}}^H}{{\bf{T}}_{ik}}{\bf{\bar q}} + {\left| {{d_{ik}}} \right|^2} = {\rm{tr}}\left( {{{\bf{T}}_{ik}}{\bf{\bar Q}}} \right){\rm{ + }}{\left| {{d_{ik}}} \right|^2},
\end{equation}
where ${\bf{\bar Q}} = {\bf{\bar q}}{{\bf{\bar q}}^H}$ and ${\rm{rank}}\left( {{\bf{\bar Q}}} \right) = 1$. Then the problem (P4.2) can be converted to the problem (P4.2.1), which can be given by
\begin{subequations}
	\begin{align}
		\left( {{\textrm{P4.2.1}}} \right){\rm{~~}}&{\rm{find~~}}{\bf{\bar Q}}, \\
		\rm{s.t.}\qquad & \begin{array}{l}
			{\rm{ }}{p_k}{\rm{tr}}\left( {{{\bf{T}}_{kk}}{\bf{\bar Q}}} \right) \!-\! {\gamma _{th}}\sum\limits_{i \ne k} {{p_i}{\rm{tr}}\left( {{{\bf{T}}_{ik}}{\bf{\bar Q}}} \right)} \! \ge\! {\gamma _{th}}\sigma _n^2{\rm{tr}}\left( {{{\bf{W}}_k}} \right)\\
			{\rm{                }} + {p_k}{\varepsilon  ^{\rm UL}_k}{\rm{tr}}\left( {{{\bf{I}}_N}{{\bf{W}}_k}} \right) - {\gamma _{th}}\sum\limits_{i \ne k} {{p_i}{\varepsilon  ^{\rm UL}_i}{\rm{tr}}\left( {{{\bf{I}}_N}{{\bf{W}}_k}} \right)} \\
			{\rm{                  + }}\left( {{\gamma _{th}}\sum\limits_{i \ne k} {{p_i} - {p_k}} } \right){\left| {{d_{ik}}} \right|^2},\forall k,
		\end{array}\\
		&{\bf{\bar Q}} \succeq 0,\\
		&{\rm{rank}}\left( {{\bf{\bar Q}}} \right) = 1,\\
		&{{{\bf{\bar Q}}}_{m,m}} = 1,m = 1,...,M + 1.
	\end{align}
\end{subequations}
Similar to the problem (P4.1.1), this problem can be rewritten as the problem (P4.2.2) as follows
\begin{subequations}
	\begin{align}
		\left( {{\textrm{P4.2.2}}} \right){\rm{~~~~}}&\min {\rm{  tr}}\left( {{\bf{\bar Q}}} \right) - \left\langle {\partial {{\left\| {{{{\bf{\bar Q}}}^{t - 1}}} \right\|}_2},{\bf{\bar Q}}} \right\rangle , \\
		\rm{s.t.}\qquad & \textrm {(57b), (57c), (57e)}.
	\end{align}
\end{subequations}
Until the stopping criterion ${\rm{tr}}\left( {{\bf{\bar Q}}} \right) - {\left\| {{\bf{\bar Q}}} \right\|_2} \le \epsilon $ is satisfied, a solution of the problem (P4.2.2) is obtained. Therefore, the robust beamforming design algorithm of IRS can be shown as $\textbf{Algorithm 2}$.	
\begin{algorithm}[H]
	\caption{The Robust Beamforming Design Algorithm of IRS by Applying DC Programming} 
	\begin{algorithmic}[1]
		\State $\textbf{Input:}$ $\tau^0 $, ${{{\bf{V}}_l}^0}$, ${{{\bf{W}}_k}^0}$, ${\bf{P}}^0$, convergence threshold $\epsilon$ and iteration index $t = 0$.
		\Repeat 
		\State For the $t$-th ineration, given $\tau^t $, ${{{\bf{V}}_l}^t}$, ${{{\bf{W}}_k}^t}$ and ${\bf{P}}^t$, solve the problem (P4.1.3) (or the problem (4.2.2)) to obtain $\bf{\bar E}^*$ (or $\bf{\bar Q}^*$). 
		\State Update $t = t + 1$.
		\Until the stopping criterion is satisfied.
		\State ${\bf{e}}^ * $ and ${\bf{q}}^ * $ can be obtained by eigenvalue decomposition of $\bf{\bar E}^*$ and $\bf{\bar Q}^*$, respectively.
		\State $\textbf{Output:}$ IRS energy beamforming vector ${\bf{e}}^ * $, information beamforming vector {${\bf{q}}^ * $}.
	\end{algorithmic}
\end{algorithm}
\subsubsection{Time allocation}
For the third sub-problem 3, we fix the HAP beamforming design $\left\{ {{{\bf{V}}_l}} \right\}$, $\left\{ {{{\bf{W}}_k}} \right\}$, ${{\bf{P}}}$ and IRS beamforming design ${\bf{e}}$, ${\bf{q}}$. The problem (P2) can be converted to the problem (P5) as follows
\begin{subequations}
	\begin{align}
		\left( {{\textrm{P5}}} \right){\rm{~~~~}}&{\rm{}}\mathop {\min }\limits_\tau  {\rm{ ~~}}\tau \sum\limits_{l = 1}^L {{\rm{tr}}\left( {{{\bf{V}}_l}} \right)}  \\
		\rm{s.t.}\qquad & 0 \le \tau  \le 1,\\
		&{\Xi ^{ - 1}}\left( {\frac{{1 - \tau }}{\tau }{p_k}} \right) \le \sum\limits_{l = 1}^L {{\rm{tr}}\left( {\left( {{{\bf{\Phi }}_k} - {\varepsilon ^{\rm DL}_k}{{\bf{I}}_N}} \right){{\bf{V}}_l}} \right)} ,\forall k.
	\end{align}
\end{subequations}
We can see that the problem (P5) is a standard LP problem and can be solved by the CVX toolbox \cite{grant2014cvx}. 
 
Finally, the three sub-problems are alternately optimized to obtain the solution of problem (P2).

\subsection{The Overall Robust Optimization Algorithm in IRS-assisted WPCN}
Based on the previous sub-problems, we propose the overall robust beamforming design and time allocation algorithm, which is summarized as $\textbf{Algorithm 3}$. In the sub-problem 1, the HAP beamforming design and user power control are determined by applying $\textbf{Algorithm 1}$. In the sub-problem 2, the IRS beaforming design is obtained through $\textbf{Algorithm 2}$. Further, time allocation in the sub-problem 3 is obtained by solving a standard LP. Finally, three sub-problems are alternately solved to achieve convergence.
\begin{algorithm}[H]
	\caption{The Overall Robust Beamforming Design and Time Allocation Algorithm} 
	\begin{algorithmic}[1]
		\State Randomly initialize $\tau^0 $, ${{\bf{V}}^{{\rm{dl}}}}$, ${{\bf{W}}^{{\rm{ul}}}}$, ${\bf{P}}^0$, ${\bf{e}}^0$ and ${\bf{q}}^0$. Initialize convergence threshold $\epsilon$ and iteration index $t = 0$.
		\Repeat
		\State Obtain HAP energy beamforming vector ${\bf{v}}_l^ * $, receiving beamforming vector ${\bf{w}}_k^ * $ and user transmit power ${p}_k^ * $ according to the $\textbf{Algorithm 1}$.
		\State Obtain IRS energy beamforming vector ${\bf{e}}^ * $, information beamforming vector {${\bf{q}}^ * $} according to the $\textbf{Algorithm 2}$.
		\State Obtain time allocation $\tau^*$ by solving an LP problem (P5). 
		\State Update $t=t+1$.
		\Until The fractional decrease of the objective value is below a threshold $\epsilon$.
		\State \Return The beamforming design and time allcation scheme.
	\end{algorithmic}
\end{algorithm}

\subsection{Computational Complexity and Convergence Analysis}
\subsubsection{Computational complexity analysis}
The complexity of $\textbf{Algorithm 1}$ mainly depends on iteratively solving the SDP problem (P3.1), (P3.2.1) and the LP problem (P3.3). In each iteration, by using the interior point method \cite{boyd2004convex}, the computational complexity of solving problem (P3.1) is ${\cal O}\left( {{N^{3.5}}} \right)$, the computational complexity of solving problem (P3.2.1) is ${\cal O}\left( {K{N^{3.5}}} \right)$, and the computational complexity of solving problem (P3.3) is ${\cal O}\left( {{K}} \right)$. Therefore, the computational complexity of $\textbf{Algorithm 1}$ is at most ${\cal O}\left( {K{N^{3.5}}} \right)$. Similarly, the computational complexity of $\textbf{Algorithm 2}$ is mainly determined by SDP problems (P4.1.3) and (P4.2.2). In each iteration, the complexity of solving problems (P4.1.3) and (P4.2.2) by the interior point method can be denoted by ${\cal O}\left( {(M+1)^{3.5}} \right)$ \cite{boyd2004convex}. While the subgradient can be computed by singular value decomposition (SVD) with complexity ${\cal O}\left( {{(M+1)^{3}}} \right)$. Therefore, the computational complexity of $\textbf{Algorithm 2}$ is at most ${\cal O}\left( {(M+1)^{3.5}} \right)$. Finally, the computational complexity of the third sub-problem is constant. In summary, let $t$ be the number of iterations required for the proposed algorithm to achieve convergence, the computational complexity of $\textbf{Algorithm 3}$ can be denoted by ${\cal O}\left( {t\left( {K{N^{3.5}} + {(M+1)^{3.5}}} \right)} \right)$.

\subsubsection{Convergence analysis}
The convergence of the proposed $\textbf{Algorithm 3}$ in IRS-assisted WPCN can be proved as follows. 

We define ${{\tau ^t}}$, ${{{\left( {{{\bf{V}}^{{\rm{dl}}}}} \right)}^t}}$, ${{{\left( {{{\bf{W}}^{{\rm{ul}}}}} \right)}^t}}$, ${{{\bf{P}}^t}}$, ${{{\bf{e}}^t}}$ and ${{{\bf{q}}^t}}$ as the $t$-th iteration solution of the problem (P3), (P4) and (P5). Herein, the objective function is denoted by ${\cal E}\left( {{\tau ^t},{{\left( {{{\bf{V}}^{{\rm{dl}}}}} \right)}^t},{{\left( {{{\bf{W}}^{{\rm{ul}}}}} \right)}^t},{{\bf{P}}^t},{{\bf{e}}^t},{{\bf{q}}^t}} \right)$. In the step 3 of $\textbf{Algorithm 3}$, since the HAP beamforming design and user power allocation can be obtained for given ${{\tau ^t}}$, ${{{\bf{e}}^t}}$ and ${{{\bf{q}}^t}}$. Hence, we have 
\begin{equation}
	{\cal E}\left( {{\tau ^t},{{\left( {{{\bf{V}}^{{\rm{dl}}}}} \right)}^t},{{\left( {{{\bf{W}}^{{\rm{ul}}}}} \right)}^t},{{\bf{P}}^t},{{\bf{e}}^t},{{\bf{q}}^t}} \right) \ge {\cal E}\left( {{\tau ^t},{{\left( {{{\bf{V}}^{{\rm{dl}}}}} \right)}^{t + 1}},{{\left( {{{\bf{W}}^{{\rm{ul}}}}} \right)}^{t + 1}},{{\bf{P}}^{t + 1}},{{\bf{e}}^t},{{\bf{q}}^t}} \right).
\end{equation}
In the step 4 of $\textbf{Algorithm 3}$, the IRS beaforming design scheme can be obtained when ${{\tau ^t}}$, ${{{\left( {{{\bf{V}}^{{\rm{dl}}}}} \right)}^t}}$, ${{{\left( {{{\bf{W}}^{{\rm{ul}}}}} \right)}^t}}$ and ${{{\bf{P}}^t}}$ are given. Herein, we also have
\begin{equation}
	{\cal E}\left( {{\tau ^t},{{\left( {{{\bf{V}}^{{\rm{dl}}}}} \right)}^{t + 1}},{{\left( {{{\bf{W}}^{{\rm{ul}}}}} \right)}^{t + 1}},{{\bf{P}}^{t + 1}},{{\bf{e}}^t},{{\bf{q}}^t}} \right) = {\cal E}\left( {{\tau ^t},{{\left( {{{\bf{V}}^{{\rm{dl}}}}} \right)}^{t + 1}},{{\left( {{{\bf{W}}^{{\rm{ul}}}}} \right)}^{t + 1}},{{\bf{P}}^{t + 1}},{{\bf{e}}^{t + 1}},{{\bf{q}}^{t + 1}}} \right).
\end{equation}Finally, in the step 5 of $\textbf{Algorithm 3}$, time allocation can be obtained when ${{{\left( {{{\bf{V}}^{{\rm{dl}}}}} \right)}^t}}$, ${{{\left( {{{\bf{W}}^{{\rm{ul}}}}} \right)}^t}}$, ${{{\bf{P}}^t}}$, ${{{\bf{e}}^t}}$ and ${{{\bf{q}}^t}}$ are fixed. Therefore, we have
\begin{equation}
	{\cal E}\left( {{\tau ^t},{{\left( {{{\bf{V}}^{{\rm{dl}}}}} \right)}^{t + 1}},{{\left( {{{\bf{W}}^{{\rm{ul}}}}} \right)}^{t + 1}},{{\bf{P}}^{t + 1}},{{\bf{e}}^{t + 1}},{{\bf{q}}^{t + 1}}} \right) \ge {\cal E}\left( {{\tau ^{t + 1}},{{\left( {{{\bf{V}}^{{\rm{dl}}}}} \right)}^{t + 1}},{{\left( {{{\bf{W}}^{{\rm{ul}}}}} \right)}^{t + 1}},{{\bf{P}}^{t + 1}},{{\bf{e}}^{t + 1}},{{\bf{q}}^{t + 1}}} \right).
\end{equation}
Based on the above, we can obtain
\begin{equation}
	{\cal E}\left( {{\tau ^t},{{\left( {{{\bf{V}}^{{\rm{dl}}}}} \right)}^t},{{\left( {{{\bf{W}}^{{\rm{ul}}}}} \right)}^t},{{\bf{P}}^t},{{\bf{e}}^t},{{\bf{q}}^t}} \right) \ge {\cal E}\left( {{\tau ^{t + 1}},{{\left( {{{\bf{V}}^{{\rm{dl}}}}} \right)}^{t + 1}},{{\left( {{{\bf{W}}^{{\rm{ul}}}}} \right)}^{t + 1}},{{\bf{P}}^{t + 1}},{{\bf{e}}^{t + 1}},{{\bf{q}}^{t + 1}}} \right).
\end{equation}
It shows that the value of the objective function after each iteration of $\textbf{Algorithm 3}$ is non-increasing. At the same time, the objective function value of the problem (P2) has a lower bound, so the convergence of $\textbf{Algorithm 3}$ can be guaranteed.

\section{Numerical results}
In this section, we demonstrate on the effectiveness of the proposed robust beamforming design and time allocation algorithm in IRS-assisted WPCN through numerical simulations. We consider a three-dimensional coordinate system in this paper, where the location of HAP is (0m, 0m, 15m), the location of IRS is (50m, 50m, 15m), and in the circle whose origin is (50m, 45m, 0m) and radius is 5m, $K=4$ users are randomly distributed. Moreover, HAP is equipped with $N=6$ antennas, and IRS is equipped with $M=20$ reflection elements. We set the antenna spacing to be half of the carrier wavelength. Meanwhile, we make the parameters of all users consistent. i.e., ${\xi _k} = 24{\rm{mW}}$, ${a_k} = 150$ and ${b_k} = 0.024$. In addition, we set $P_{\max}=43$dBm, $\sigma ^2=-70$dBm and ${\gamma}_{th} =10$dB in our numerical simulations. The path loss exponents are set as $\alpha=3$, $\beta=2.2$ and $o=2.5$. The path loss with a reference distance of 1m is set to $C_0=-30$dB. We set Rician factor $\kappa$ to 3dB, and we set the convergence threshold of the proposed algorithm to $10^{-3}$. 

First, we verify the convergence of the proposed robust beamforming design and time allocation algorithm in IRS-assisted WPCN. Fig. 3 shows the transmit energy of HAP varies with the number of iterations under different IRS reflection elements. We can see that the transmit energy gradually decreases as the number of iterations increases. The proposed algorithm can quickly achieve convergence and has good convergence performance. In addition, we compare the impact of different IRS reflection element numbers on system performance. Specifically, we respectively compare the performance of the proposed algorithm when the number of IRS reflection elements are 20, 40, and 60. It can be found that the larger the number of IRS reflection elements, the lower the HAP transmit energy. This also illustrates the importance of IRS for WPCN, which can reduce the transmit energy of the HAP by increasing the number of reflection elements.

\begin{figure}[htbp]
	\begin{minipage}[t]{0.5\linewidth}
		\centering
		\includegraphics[width=8cm]{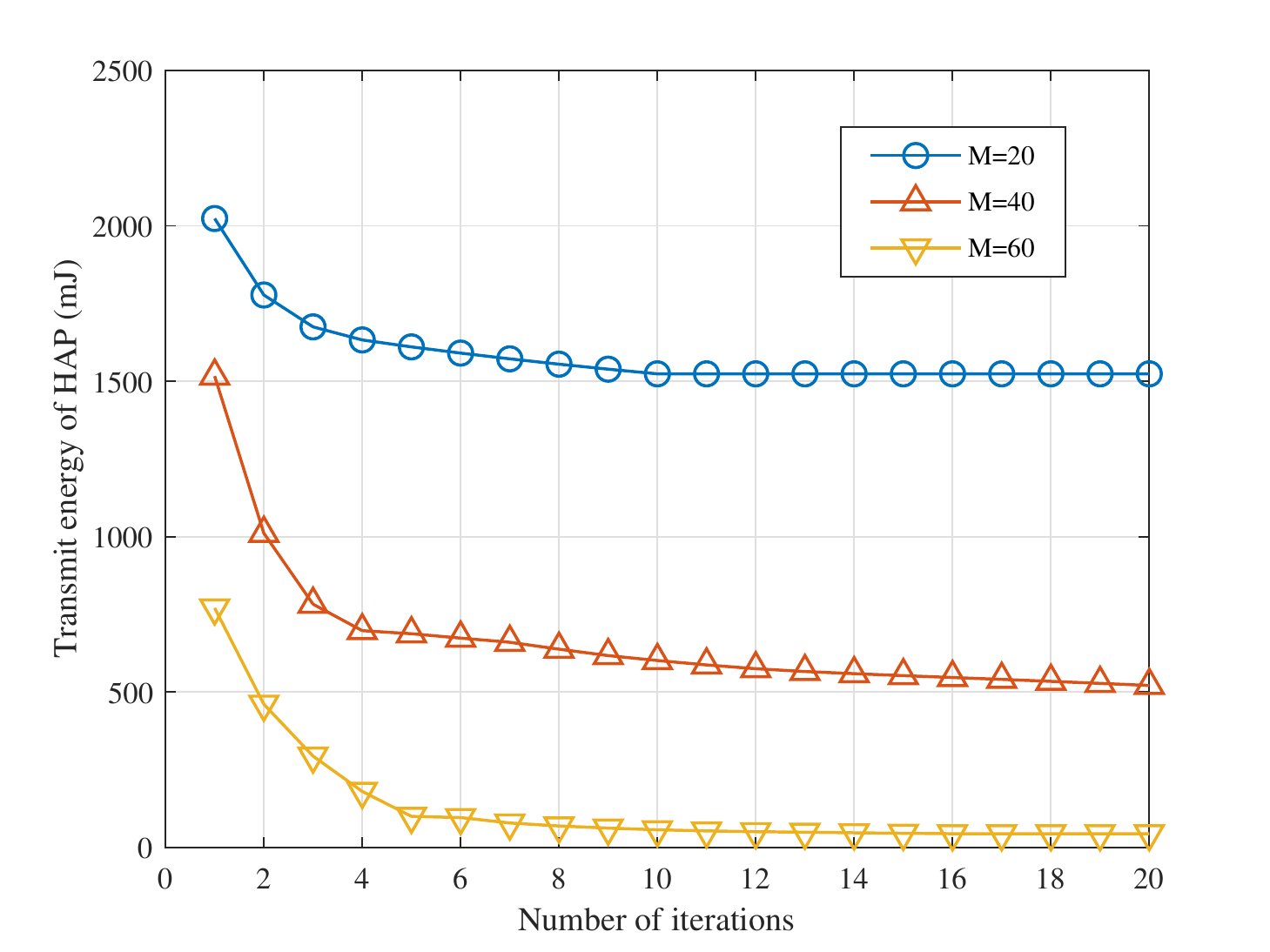}
		\caption{The convergence of the proposed robust \protect\\beamforming design and time allocation algorithm.}
	\end{minipage}%
	\begin{minipage}[t]{0.5\linewidth}
		\centering
		\includegraphics[width=8cm]{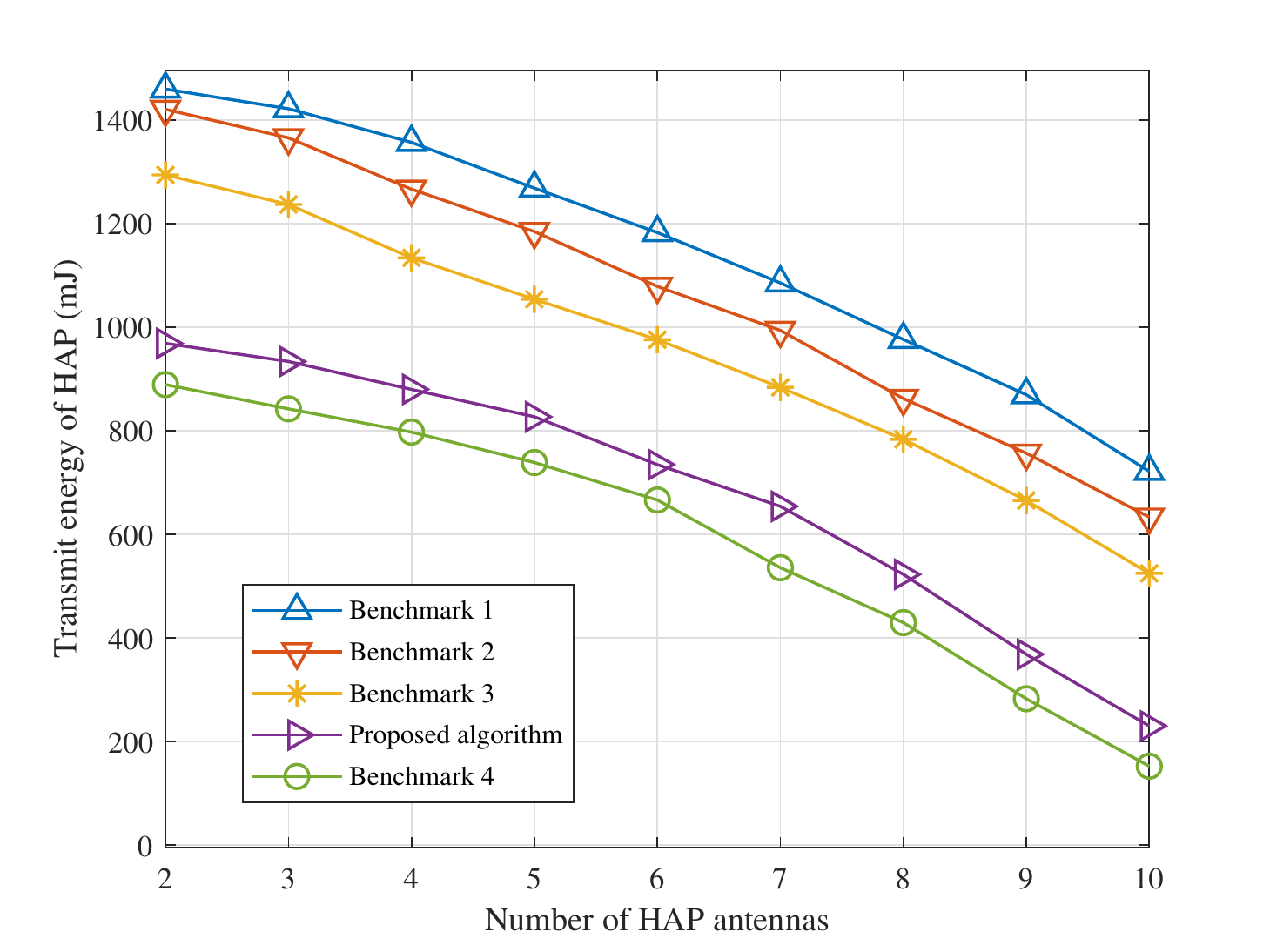}
		\caption{The HAP transmit energy versus the number \protect\\of HAP antennas.}
	\end{minipage}
\end{figure}

In this paper, we demonstrate the performance of the proposed robust beamforming design and time allocation algorithm in IRS-assisted WPCN by comparing with other benchmarks. (1) $\textbf{benchmark 1}$ (i.e., without-IRS): In this case, the IRS is no longer deployed, so there is no need to consider the combined channel, that is, the solution of the second sub-problem is not considered. (2) $\textbf{benchmark 2}$ (i.e., IRS-random-phase): In this case, we deploy IRS, but do not optimize its phase, and adopt a random phase. (3) $\textbf{benchmark 3}$ (i.e., IRS-phase-SDR): In this case, we deploy IRS and optimize its phase, but applying the SDR approach. (4) $\textbf{benchmark 4}$ (i.e., perfect-CSI): In this case, we consider beamforming design and time allocation with perfect CSI in IRS-assisted WPCN.

Next, we investigate the behavior of the HAP transmit energy with the number of HAP antennas. We can see that from Fig. 4 that the HAP transmit energy under different benchmarks decreases as the number of the HAP antennas increases. This is because that the more antennas of the HAP, the stronger the spatial diversity gain, and the lower the required transmit energy. This also stimulate our motivation to introduce large-scale antenna systems. In addition, it can be seen that in the case of the same number of the HAP antennas, the performance of our proposed algorithm is superior to benchmark 1 and benchmark 2, which reflects the superiority of IRS-assisted WPCN and the necessity of optimizing its phase. Meanwhile, its performance is better than that of benchmark 3.  The main reason is that after relaxing the rank-one constraint, the solution of the problem obtained by Gaussian randomization may not satisfy the rank-one constraint, i.e., the relaxation is not tight. And the performance of the algorithm will decrease as the problem size increases. The DC algorithm can solve the problem without dropping the non-convex rank-one constraint. Therefore, the performance of the DC algorithm is better than that of the Gaussian randomization algorithm. Compared with benchmark 4, although the performance of our proposed algorithm is slightly inferior, the perfect CSI of the practical system is almost impossible. Therefore, the robustness of our proposed algorithm is better, and it is closer to the deployment of the practical system.

Fig. 5 shows the variation of the HAP transmit energy with the number of IRS reflection elements. Under different benchmarks, HAP transmit energy continues to decrease as the number of IRS reflection elements increases. This is owing to as the number of IRS elements increases, the number of combined channels and the channel gain increase, so that the HAP transmit energy decreases. Meanwhile, this also shows the advantages of IRS. With the assistance of IRS, the performance of WPCN in IoT networks can be improved. In addition, we also found that when the number of IRS reflection elements is the same, the performance of our proposed algorithm still outperforms benchmarks 2 and 3, and slightly inferior to 4. The main reason is that benchmark 2 does not optimize the phase of the IRS. As the number of reflection elements increases, the performance of SDR becomes weaker. Therefore, the performance of our proposed algorithm outperforms that of benchmark 3. This also reflects the advantages of the DC algorithm we used in solving the second sub-problem. Although the performance of our proposed algorithm is slightly inferior to benchmark 4, it is robust and more conducive to practical deployment.

Fig. 6 illustrates the variation of the HAP transmit energy with the number of users. From the Fig. 6, we can find that the transmit energy of the HAP increases as the number of users increases under different benchmarks. This is mainly because as the number of users in the HAP coverage area increases, it needs to provide more energy to meet the user's quality-of-service (QoS) requirements and energy harvesting requirements. In addition, when the number of users is the same, the performance of our proposed algorithm still outperforms benchmark 1, 2 and 3, and slightly inferior to benchmark 4 under ideal conditions. The reasons are similar to those above-mentioned .

\begin{figure}[htbp]
	\begin{minipage}[t]{0.5\linewidth}
		\centering
		\includegraphics[width=8cm]{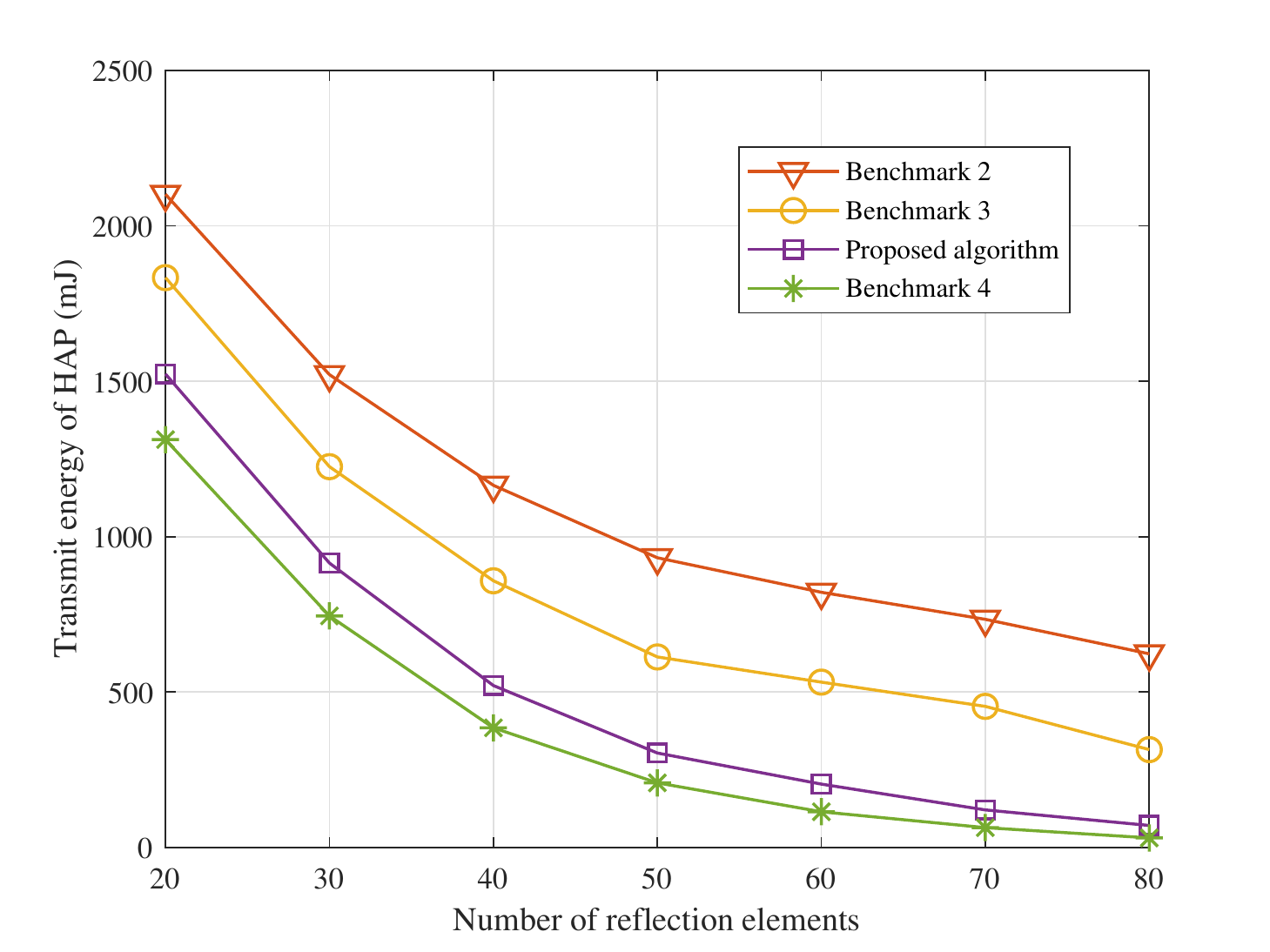}
		\caption{The HAP transmit energy versus the number of\protect\\ reflection elements.}
	\end{minipage}%
	\begin{minipage}[t]{0.5\linewidth}
		\centering
		\includegraphics[width=8cm]{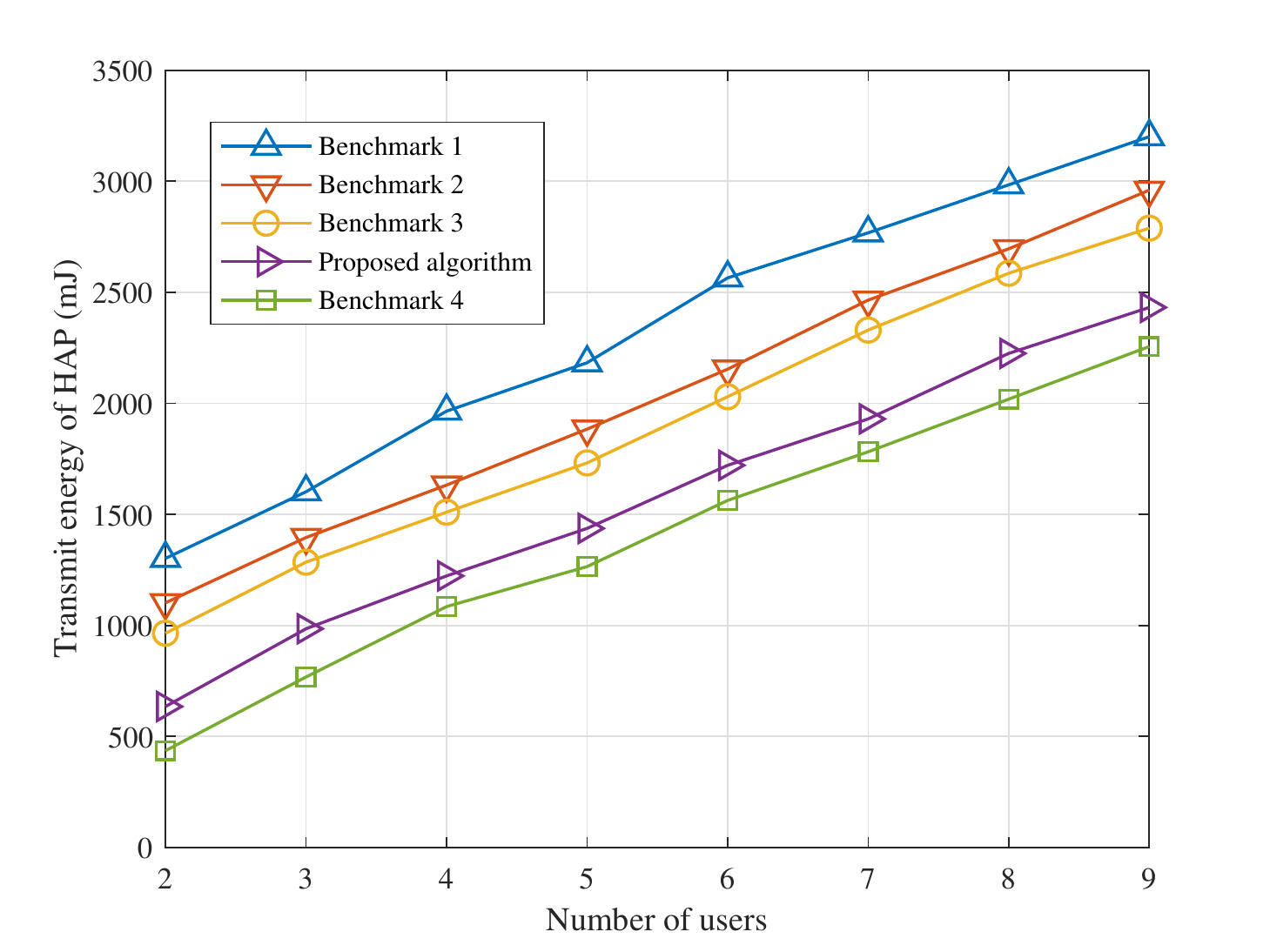}
		\caption{The HAP transmit energy versus the number of users.}
	\end{minipage}
\end{figure}

Fig. 7 shows the variation of the HAP transmit energy with the user's SINR target under different benchmarks. When the user's SINR target increases, the transmit energy of the HAP increases. In order to meet the user's higher rate requirements or higher QoS requirements, the HAP needs to allocate more energy to users so that they have greater transmit energy in the UL, so the transmit energy of the HAP needs to be continuously increased. In addition, when the user's SINR target is the same, the performance comparison of several benchmarks and the proposed beamforming design and time allocation algorithm in IRS-assisted WPCN still meets the relationship explained above.

In order to illustrate the robust beamforming design we proposed, Fig. 8 depicts the HAP transmit energy of the proposed algorithm versus the spectral norm of different channel error matrices. Herein, we set $\Delta {\bf{E}}_k^{{\rm{DL}}} = \Delta {\bf{E}}_k^{{\rm{UL}}}$. The abscissa of Fig. 8 is logarithm. Under different benchmarks, as the spectral norm of the error matrix increases, the transmit energy of the HAP continues to increase. This is because that in the problem (P2), we consider the worst-case, i.e., to meet the worst-case user's QoS requirements and energy harvesting requirements. Therefore, the greater the channel error, the greater the transmit energy required by the HAP to meet the the worst-performing user. Since it is impossible to obtain perfect CSI at the HAP in an practical system, the beamforming design considering a certain channel error is robust and is more conducive to deployment in the practical system. In addition, through the above numerical simulations, we can also find that the performance gap between our proposed algorithm and perfect CSI under different conditions is not very large but it is better than other benchmarks, achieving a good trade-off between practical and theoretical.
\begin{figure}[htbp]
	\begin{minipage}[t]{0.5\linewidth}
		\centering
		\includegraphics[width=8cm]{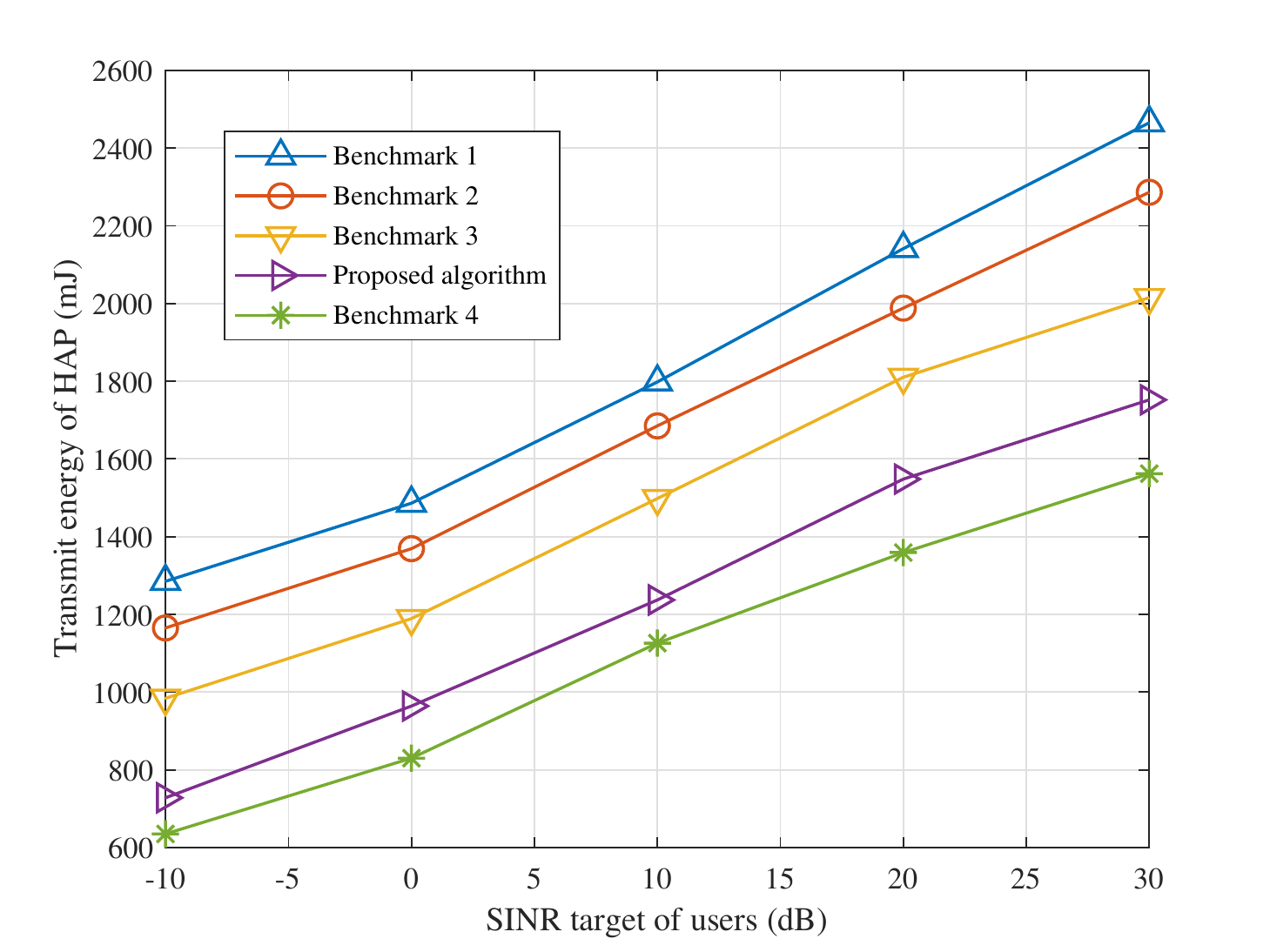}
		\caption{The HAP transmit energy versus the SINR target\protect\\ of users.}
	\end{minipage}%
	\begin{minipage}[t]{0.5\linewidth}
		\centering
		\includegraphics[width=8cm]{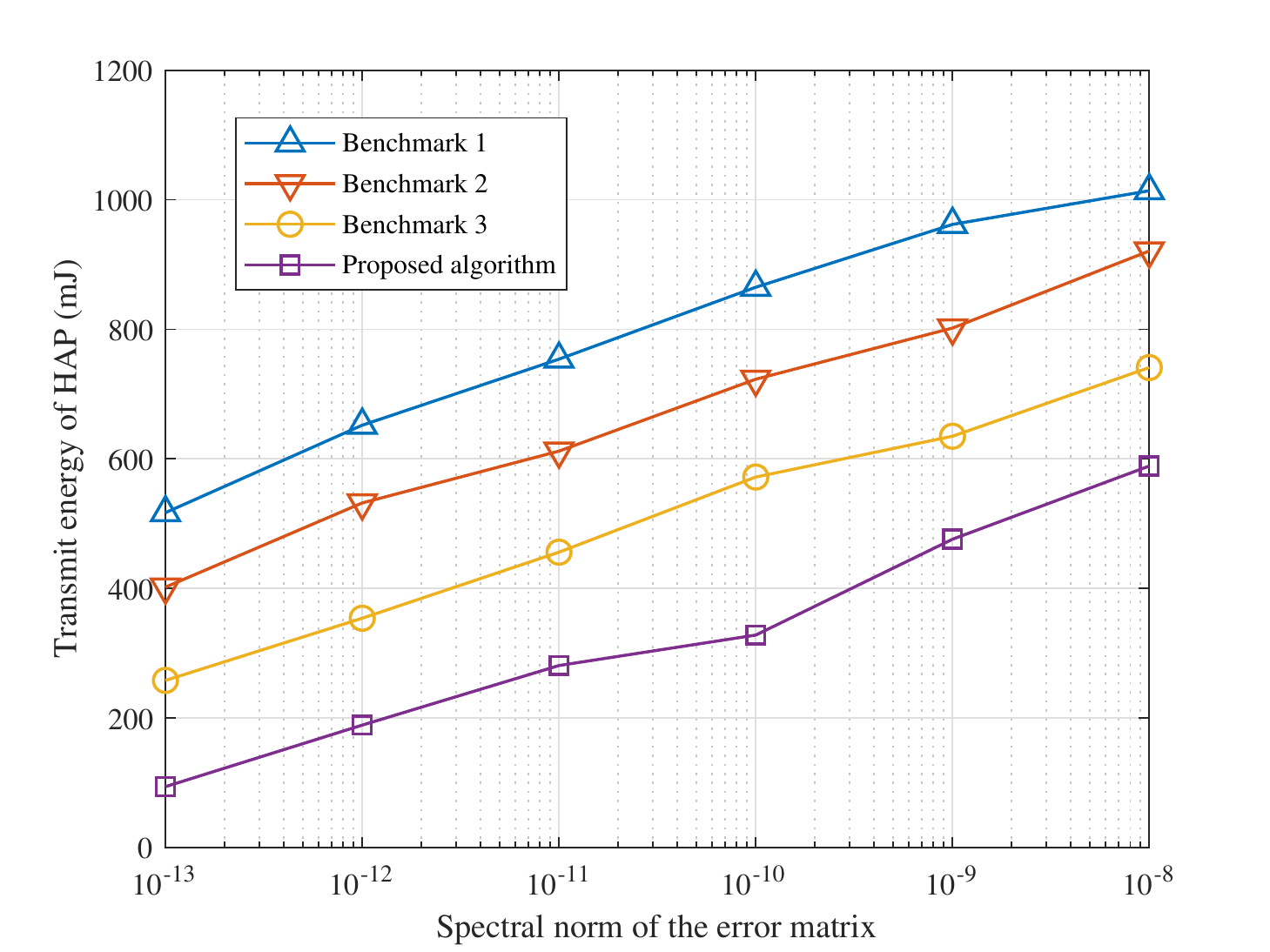}
		\caption{The HAP transmit energy versus the spectral norm \protect\\of the error matrix.}
	\end{minipage}
\end{figure}

Finally, we investigate the impact of IRS location on system performance. Herein, we let the positions of HAP and IRS be (0m, 0m, 0m) and ($x_{irs}$, 0m, 0m), respectively. Users are randomly distributed in a circle with a center of (60m, 0m, 0m) and a radius of 10m, so the distance between the center of user distribution and the HAP is $d_{htu}=60$m. In addition, we set the path loss coefficient $\alpha  = \beta  = 2.5$. The position of the IRS changes from 10m to 50m with a step size of 10m. As shown in Fig. 9, for benchmark 1, it has no IRS assistance, so the HAP transmit energy does not change with the position of the IRS. For benchmark 2, 3, 4 and our proposed algorithm, the transmit energy of HAP first increases and then decreases with the increase of $x_{irs}$. When $x_{irs}$ is about 30m, the transmit energy of HAP is the largest. In order to simplify the analysis, we only consider the large-scale channel gain related to the distance, which can be expressed as
\begin{equation}
	\varpi \left( d \right) = \frac{{{{10}^{ - 3}}}}{{{{\left( {{d_{htu}}} \right)}^{2.5}}}} + \frac{{{{10}^{ - 6}}}}{{{{\left( {{d_{htr}}{d_{rtu}}} \right)}^{2.5}}}},
\end{equation}
where ${d_{htr}}$ denotes the distance from the HAP to the IRS, ${d_{rtu}}$ represents the distance from the IRS to the user's distribution center, and ${d_{htu}} = {d_{htr}} + {d_{rtu}}$. Therefore, when ${d_{htr}} = {d_{rtu}} = \frac{{{d_{htu}}}}{2}$, the channel gain is the smallest, so the required HAP transmit energy is the largest.
\begin{figure}
	\centerline{\includegraphics[width=8cm]{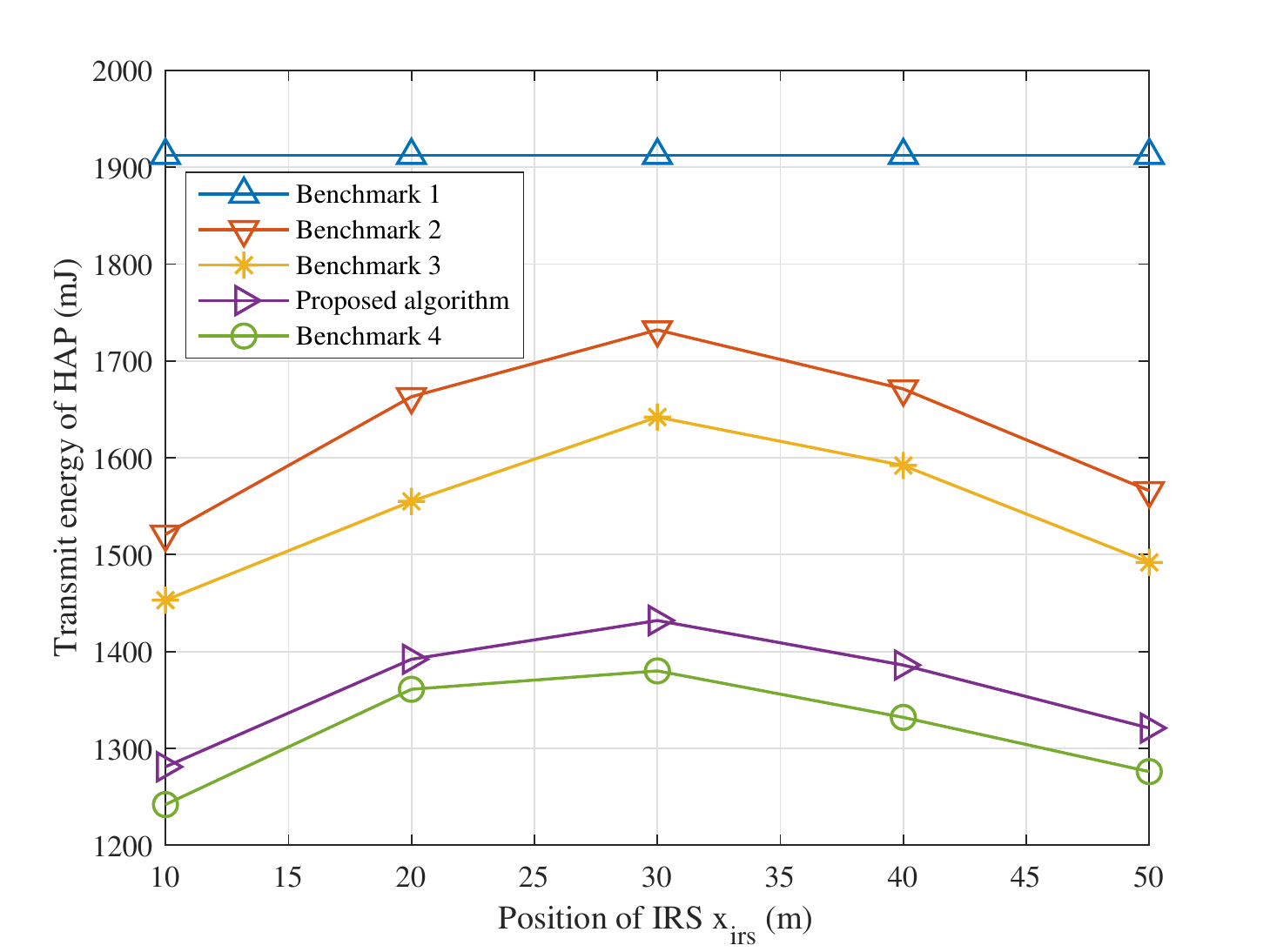}}
	\caption{The HAP transmit energy versus the position of IRS.}
\end{figure}
\section{Conclusions}
This paper investigates the HAP transmit energy minimization problem for the IRS-assisted WPCN. Specifically, time allocation, HAP energy beamforming, receiving beamforming, user transmit power allocation, IRS energy reflection coefficient and information reflection coefficient are jointly optimized by applying AO framework with DC programming. The problem is first transformed and then divided into three sub-problems. In the first sub-problem, the HAP energy beamforming, receiving beamforming and user transmit power allocation are optimized by solving SDP problems and a LP problem. Next, in the second sub-problem, IRS energy reflection coefficient and information reflection coefficient are obtained by applying variable substitution and DC programming. DC programming is introduced to deal with non-convex rank-one constraints, which has better performance than SDR. In the third sub-problem, the time allocation scheme is obtained by solving an LP problem. Finally, the three sub-problems are solved alternately to achieve convergence. In addition, the computational complexity and convergence of the proposed robust beamforming design and time allocation algorithm are analyzed. Numerical simulation results show that the performance of our proposed algorithm is better than other benchmarks, and the auxiliary role of IRS is extremely important, which can greatly decrease the HAP transmit energy with low cost in practice.

\ifCLASSOPTIONcaptionsoff
  \newpage
\fi



\bibliographystyle{IEEEtran}
\bibliography{reference}
\end{document}